\documentclass[12pt]{article}
\usepackage{latexsym,epsfig,amssymb,euscript,amsmath,verbatim,subfigure,lipsum}
\usepackage{cite}
\newcommand{\beq}{\begin{equation}}
\newcommand{\eeq}{\end{equation}}
\newcommand{\bea}{\begin{eqnarray}}
\newcommand{\eea}{\end{eqnarray}}
\newcommand{\Fslash}{{\not\! F}}

\newcommand{\misset}{\slash\!\!\!\!E_T}

\newcommand{\MET}{E\llap{/\kern1.5pt}_T}

\begin{document}

\pagestyle{empty}

\begin{flushright}
IPPP/13/98\\
DCPT/13/196
\end{flushright}

\begin{center}

{\LARGE{\bf Multiphoton signatures of goldstini at the LHC}}


\vspace{0.8cm}

{\large{ Gabriele Ferretti$^{1}$, Alberto Mariotti$^{2}$,\\ 
Kentarou Mawatari$^{3,5}$ and Christoffer Petersson$^{1,4,5}$\\[5mm]}}

{\small{
{}$^1$  Department of Fundamental Physics,\\
Chalmers University of Technology, 412 96 G\"oteborg, Sweden\\
{}$^2$ Institute for Particle Physics Phenomenology,\\
Department of Physics, Durham University, DH1 3LE, United Kingdom\\
{}$^3$ Theoretische Natuurkunde and IIHE/ELEM,\\
Vrije Universiteit Brussel, Pleinlaan 2, 1050 Brussels, Belgium\\
{}$^4$ Physique Th\'eorique et Math\'ematique,\\
Universit\'e Libre de Bruxelles, C.P. 231, 1050 Brussels, Belgium\\
{}$^5$ International Solvay Institutes, Brussels, Belgium}}

\vspace{0.7cm}

{\bf Abstract}

\end{center}


\noindent We study models of gauge mediated SUSY breaking with more than
one hidden sector. In these models the neutralino sector of the MSSM is
supplemented with additional light neutral fermions, the nearly massless gravitino and the massive pseudo-goldstini. For the case where the Bino is the lightest ordinary SUSY particle, its preferred decay is to a photon and the heaviest pseudo-goldstino,
 which generically cascades down to lighter pseudo-goldstini,
 or to the gravitino, in association with photons. This gives rise
to multiphoton plus missing energy signatures at the LHC. We investigate
in detail simplified models where the SUSY spectrum consists of the
right-handed sleptons, a Bino-like neutralino, the pseudo-goldstini and the
gravitino. We compare against existing LHC searches and show that the
sensitivity to our models could be significantly improved by relaxing
the kinematic cuts and requiring additional final state particles. We propose inclusive searches in the final states $({\geqslant}3)\gamma+\MET$ and
$\ell^{+}\ell^{-}+({\geqslant}2)\gamma+\MET$, the former being sensitive
to any production mode and the latter being optimized for slepton pair
production.
We show that they could lead to an observation (or strong constraints)
already with the data set from LHC Run I, and present prospects for LHC Run II.

\newpage
\tableofcontents

\setcounter{page}{1}
\pagestyle{plain}
\renewcommand{\thefootnote}{\arabic{footnote}}
\setcounter{footnote}{0}

\section{Introduction}

The absence of any signal of supersymmetry (SUSY)
at the LHC after \mbox{Run I} motivates studies of 
extensions of the minimal supersymmetric standard model (MSSM), as well as non-standard experimental searches. As
an example of this, we investigate how the standard phenomenology of
models with gauge mediated supersymmetry breaking (GMSB) can be
significantly modified by the assumption that
SUSY is broken in more than one hidden sector. Each of
the hidden sectors provides a neutral fermion, with one linear
combination corresponding to the goldstino (GLD), which gets eaten by
the gravitino becoming its longitudinal components. The other linear
combinations correspond to the so called pseudo-goldstini (PGLDs),
which acquire masses both at the tree and loop level but are in general
lighter than the lightest ordinary supersymmetric particle (LOSP) of the
MSSM.\footnote{In standard GMSB with only one hidden sector, the LOSP is
synonymous to the NLSP.} In contrast to standard GMSB models, where the
next-to-lightest superpartner (NLSP) always decays to its SM partner
and the LSP gravitino, the LOSP may decay to the heaviest PGLD, which
subsequently decays to the second heaviest PGLD and so on. In each
step of the decay chain, SM particles are emitted. Hence, while the
presence of PGLDs generically make the final state spectrum softer, their presence
can increase the number of final state particles and open up new search
channels.

Models with multiple hidden sectors have been investigated in several
papers. In the context of gravity mediation, they have been discussed
in~\cite{Cheung:2010mc}; see
also~\cite{Benakli:2007zza,
Cheng:2010mw}.
The first study of multiple hidden sectors in the context of GMSB was
done in~\cite{Argurio:2011hs}. The collider phenomenology of GMSB with
goldstini was discussed in~\cite{Argurio:2011gu} for the case where the
LOSP was a gaugino-like neutralino or a stau, and in~\cite{Liu:2013sx}
for the case of higgsino LOSP. Some further remarks relevant to these
models can be found
in~\cite{McGarrie:2012np
}.
Note that in all these previous investigations the attention was focused mainly
on the case of two SUSY breaking sectors.

It turns out that models with more than two hidden sectors are
qualitatively different from the two sector case. Consider for
definitiveness the case of three hidden sectors and denote the heaviest
PGLD by $\tilde{G}''$, the second heaviest by $\tilde{G'}$, and the
nearly massless LSP gravitino by $\tilde G$.
 We assume R-parity conservation and consider the
case where the LOSP is a Bino-like neutralino $\tilde{\chi}$.
We will
show that $\tilde{\chi}$ dominantly decays promptly to a photon and the heaviest
$\tilde{G}''$. Moreover, since the coupling of $\tilde{G}''$ to $\tilde G'$ can be significantly stronger than the coupling to $\tilde G$, the dominant decay of $\tilde{G}''$ is to $\tilde G'$. 
Among the decay modes of $\tilde{G}''$ we
find that the leading one is to a photon and $\tilde{G}'$,
which can be prompt. 
On the other hand, the
$\tilde{G}'$ decay to the gravitino
$\tilde{G}$ typically takes place outside the detector. Therefore, the
cascade decay of $\tilde\chi$ leads to $2\gamma+\tilde{G}'$,
where the (collider stable) $\tilde{G}'$ carries away $\MET$. From the
collider point of view, $\tilde{G}$ plays a minor role in these models,
except for the particular case where the mass spectrum is very squeezed.

Depending on the way in which the neutralino LOSPs are pair produced,
the multiphotons in the final state will be accompanied with different
final state particles, such as jets or leptons in the case of colored or
electroweak production, respectively. We focus on a simplified model where the dominant
production mode is pair production of right-handed sleptons, each of
which decays to a lepton and a Bino-like neutralino. This model is
motivated by the fact that, since the Bino and the right-handed sleptons
interact only via the SM hypercharge, they are generically the lightest
superpartners in GMSB. Further motivation comes from the fact that, in
order to accommodate a 125 GeV Higgs boson, the colored states in GMSB,
in particular the stops and gluinos, are typically pushed up in the multi-TeV
range (see e.g. \cite{Draper:2011aa
}).

We investigate the process
$pp\to\tilde{\ell}^{+}_R\tilde{\ell}^{-}_R\to\ell^+\ell^-+(2/4)\gamma+\MET$
at the LHC in detail, where the number of photons 2/4 depends on
whether we consider two or three hidden sectors. These signal processes are compared against currently available LHC
searches, such as the searches for
$\gamma\gamma+\MET$~\cite{Aad:2012zza},
$\ell+\gamma+\MET$~\cite{ATLASphotonlepton} and
$\ell^+\ell^-+\MET$~\cite{CMSdileptons}.
However, due to tight cuts and/or large backgrounds, these searches turn out to have poor sensitivity to our simplified GMSB models with goldstini.

We propose two searches that would be optimized for these kind of GMSB
models with multiple hidden sectors. The first one is a fully inclusive
search in the final state $({\geqslant}3)\gamma+\MET$, which would be
sensitive to any production mode, including colored production. The
other search we propose is in the final state
$\ell^{+}\ell^{-}+({\geqslant}2)\gamma+\MET$, optimized to probe the
simplified models we consider, with slepton pair production. Since the
PGLDs are massive and since the mass splittings may be small, the
emitted photons will be rather soft. Therefore, in order to probe such
goldstini scenarios, we suggest to relax the selection cuts as
much as possible. 
The possibility of searching for standard GMSB in final
states involving leptons, photons and $\MET$ was discussed already in
some of the original studies of the experimental signatures of
GMSB~\cite{Dimopoulos:1996vz
}.

This paper is organized in the following way.
In Section~\ref{setup} we review the theoretical setup involving
multiple SUSY breaking sectors. The PGLDs and the GLD extend the MSSM
neutralino sector and we briefly discuss the structure of the
corresponding mass matrix and mixing.
In Section~\ref{simp} we introduce two simplified models of GMSB with
goldstini, a two-sector model and a three-sector model, and discuss the
neutralino LOSP and PGLD decays. 
In Section~\ref{sign} we study multiphoton signatures at the LHC for the
simplified models, and compare the signals against existing LHC searches
to constrain the models. We also propose new search strategies for such
goldstini scenarios. 
We conclude in Section~\ref{Conclusions}.

\section{Theoretical setup: Multiple SUSY breaking sectors}
\label{setup}

In the general case where the visible SUSY sector (which, for
definitiveness, we take to be the MSSM) is coupled to $n$ SUSY breaking
sectors (not directly interacting with each other), the MSSM soft masses can get contributions from all the $n$
sectors. For example, the Bino, Wino and the down/up type Higgs soft
masses, and the soft $B$-parameter can be written as
\begin{equation}
\label{soft}
 M_{B}=\sum_{i=1}^n M_{B(i)}~,\quad
 M_{W}=\sum_{i=1}^n M_{W(i)}~,\quad
 m_{H_{d/u}}^2=\sum_{i=1}^n m_{H_{d/u}(i)}^2~,\quad
 B=\sum_{i=1}^n B_{(i)}~.
\end{equation}
If the interactions between the $n$ hidden sectors and the MSSM were to be switched off (and if gravitational interactions were neglected), as a consequence of spontaneous SUSY breaking,
each of the $n$ hidden sectors would give rise to a massless goldstino. However, for non-vanishing couplings to the MSSM, this degeneracy is broken and only the linear combination corresponding to the true goldstino is protected. We denote the $n$ ``would-be"-goldstino Weyl fermions by $\tilde{\eta}_i$, $i=1,\cdots n$. They extend the usual $4\times4$ MSSM neutralino mass matrix to a $(4+n)\times(4+n)$ symmetric mass matrix $\mathcal{M}$ (assumed here to be real). In the gauge eigenbasis $(\tilde{B},\tilde{W}^{(3)},\tilde{H}_d^0,\tilde{H}_u^0,\tilde{\eta}_1,\cdots,\tilde{\eta}_n)$, $\mathcal{M}$ takes the following form
\beq
\label{M}
\mathcal{M}=
\left(\begin{array}{cc}
\mathcal{M}_{4\times 4}& \mathcal{M}_{4\times n} \\
\mathcal{M}_{n\times 4}& \mathcal{M}_{n\times n} \\
 \end{array}\right)~.
\eeq
The usual MSSM neutralino block is given by
\begin{equation}
\label{M44}
\mathcal{M}_{4\times 4}=
{\small
\left(\begin{array}{cccc}
 M_{B} & 0 & -m_Z \sin\theta_w \cos\beta & m_Z \sin\theta_w \sin\beta  \\
 0 & M_{W} & m_Z \cos\theta_w \cos\beta & -m_Z \cos\theta_w \sin\beta \\
 -m_Z \sin\theta_w \cos\beta & m_Z \cos\theta_w \cos\beta & 0 & -\mu \\
 m_Z \sin\theta_w \sin\beta & -m_Z \cos\theta_w \sin\beta &- \mu & 0  \\
 \end{array}\right)}~,
\end{equation}
\normalsize
where $m_Z^2=(g_1^2+g_2^2)v^2/2$,
$\sin\theta_w=g_1/\sqrt{g_1^2+g_2^2}$, $v_d=v\cos\beta$,
$v_u=v\sin\beta$ and $v=174$~GeV.

In order to obtain the mixing terms between the MSSM neutralinos and the $\tilde{\eta}$'s, i.e.~the $\mathcal{M}_{4\times n}$ block in~\eqref{M}, we study the following SUSY operators which give rise to the soft parameters in~\eqref{soft} (the operator giving rise to the Wino mass is analogous to the one for the Bino mass):
\begin{align}
- \int d^2\theta \frac{M_{B(i)}}{2 f_i}X_i W W & \supset
- \frac{M_{B(i)}}{2}\left( \tilde{B} \tilde{B}-\frac{\sqrt{2}}{f_i} \tilde{\eta}_i  \tilde{B}D_{Y}-\frac{i}{\sqrt{2}f_i} \tilde{B}  \sigma^{\mu}\bar{\sigma}^{\nu} \tilde{\eta}_i B_{\mu\nu} \right)~,\label{Bino}\\
- \int d^4 \theta \frac{m_{H_{d/u}(i)}^2}{f_i^2} X^\dagger_i X_i  H_{d/u}^{\dagger} H_{d/u}  & \supset
- m_{H_{d/u}(i)}^2 \left( h_{d/u}^{0\,\dagger}h_{d/u}^0 -\left(\frac{1}{f_i}\tilde{\eta}_i \tilde{H}_{d/u}^{0} h_{d/u}^{0\,\ast} +\mathrm{c.c.}\right)\right)~,\label{Hsoft}\\
- \int d^2 \theta \frac{B_{(i)}}{f_i} X_i  H_d H_u & \supset
- B_{(i)}\left( h_d^0 \,h_u^0 -\frac{1}{f_i} \tilde{\eta}_i \left(\tilde{H}_d^0 h_u^0 + \tilde{H}_u^0 h_d^0\right) \right)~,\label{Bmu}
\end{align}
where the $f_i$'s are the SUSY breaking scales of the different sectors and where the fermions $\tilde{\eta}_i$ resides in the associated non-linear chiral superfields \cite{Samuel:1982uh
},
\begin{equation}
\label{nl}
X_{i}=\frac{\tilde{\eta}_{i}^{\,2}}{2F_{i}}+\sqrt{2}\theta \,\tilde{\eta}_{i} +\theta^2 F_{i}
\end{equation}
with $\langle F_i \rangle=f_i$. While the first term on the RHS of each of the Eq.~\eqref{Bino}, \eqref{Hsoft} and \eqref{Bmu} provide the corresponding soft terms in~\eqref{soft}, the second terms, upon inserting the vacuum expectation values (VEVs) for the auxiliary $D_Y$-term and the Higgs scalars, give rise to mixings between the MSSM neutralinos and the $\tilde{\eta}$'s, namely
\beq
\label{offdiag}
\mathcal{M}_{4\times n}=
{\large
\left(\begin{array}{ccc}
 -\frac{M_{B(1)}\langle D_{Y}\rangle}{\sqrt{2}f_1} & \cdots & -\frac{M_{B(n)}\langle D_{Y}\rangle}{\sqrt{2}f_n}  \\
 -\frac{M_{W(1)}\langle D_{T^3}\rangle}{\sqrt{2}f_1}& \cdots & -\frac{M_{W(n)}\langle D_{T^3}\rangle}{\sqrt{2}f_n} \\
 -\frac{ v\left(m_{H_d(1)}^2 \cos\beta +B_{(1)}\sin\beta \right)}{f_1}& \cdots &  -\frac{ v\left(m_{H_d(n)}^2 \cos\beta +B_{(n)}\sin\beta \right)}{f_n} \\
 -\frac{ v\left(m_{H_u(1)}^2 \sin\beta +B_{(1)}\cos\beta \right)}{f_1}& \cdots &  -\frac{ v\left(m_{H_u(n)}^2 \sin\beta +B_{(n)}\cos\beta \right)}{f_n}  \\
 \end{array}\right)~,
 }
\eeq
where $\langle D_{Y} \rangle=- g_1 v^2 \cos 2\beta/2$ and $\langle D_{T^3} \rangle=g_2 v^2 \cos 2\beta/2$.

Concerning the mass matrix block $\mathcal{M}_{n\times n}$ in~\eqref{M}, there are two ways in which the $\tilde{\eta}$'s acquire masses. First, from the SUSY operators in Eqs.~\eqref{Bino}, \eqref{Hsoft} and \eqref{Bmu}, one sees that the SUSY breaking in the visible sector generates diagonal tree level mass terms for the $\tilde{\eta}$'s~\cite{Argurio:2011hs}. For example, by taking the (VEV of the) auxiliary $D_Y^2$-term from $WW$ of the SUSY operator in~\eqref{Bino}, one obtains a mass term for $\tilde{\eta}_i\tilde{\eta}_i$ (i.e.~the lowest component of $X_i$ in~\eqref{nl}), with mass matrix entry given by $M_{B(i)} \langle D_{Y}\rangle^2/(2f_i^2)$.\footnote{Note that the off-diagonal mixing term between the Bino and the $\tilde{\eta}_i$ in the top row of~\eqref{offdiag} assures the presence of a zero eigenvalue, corresponding to the goldstino.}
However, since this, as well as the other tree level contributions to the diagonal $\tilde{\eta}$ mass terms, are suppressed by $1/f_i^2$, these contributions are in general negligible.

The second way in which the $\tilde{\eta}$'s acquire masses is through radiative corrections. Even though the different SUSY breaking sectors do not talk to each other at tree level, they can interact at the loop level. In~\cite{Argurio:2011hs} it was shown that the leading contributions to $\mathcal{M}_{n\times n}$ in \eqref{M} are obtained by using the SUSY operators in Eqs.~\eqref{Bino}, \eqref{Hsoft} and \eqref{Bmu} and integrating out the gauge and Higgs superfields at one loop.
The precise contributions to the elements in $\mathcal{M}_{n\times n}$ in~\eqref{M} is strongly model-dependent and can only be obtained by specifying the dynamics of the hidden/messenger sectors and computing the two point functions $\langle \tilde \eta_i \tilde \eta_j \rangle$. However, by using the fact that SUSY is spontaneously broken we can say something general about the structure of $\mathcal{M}_{n\times n}$.

The full $(4+n)\times (4+n)$ mass matrix $\mathcal{M}$ in~\eqref{M} will have a zero eigenstate corresponding to the true goldstino (GLD) $\tilde{G}$. Since the $f_i$'s are taken to be much greater than the VEVs of the auxiliary $D$ and $F$ terms of the gauge and Higgs superfields, the true goldstino will, to a good approximation, be aligned with the linear combination involving only the $\tilde{\eta}$'s, i.e.~$f_1 \tilde{\eta}_1 + \dots + f_n \tilde{\eta}_n$. The remaining $(n-1)$ eigenstates correspond to the pseudo-goldstinos (PGLDs) $\tilde{G}^{(a)}$, $a=1,\dots, n-1$.

The fact that the linear combination $f_1 \tilde{\eta}_1 + \dots + f_n \tilde{\eta}_n$ forms a zero eigenvector of $\mathcal{M}_{n\times n}$ imposes $n$ conditions on the matrix $\mathcal{M}_{n\times n}$. Hence, we can express the diagonal entries of $\mathcal{M}_{n\times n}$ in terms of the off-diagonal entries $\mathcal{M}_{ij}$ ($i{<}j$), which are model-dependent unknown parameters,
\beq
\mathcal{M}_{n \times n} {=}
{\small
\left(
\begin{array}{cccc}
 -\frac{f_2 \mathcal{M}_{12}+f_3 \mathcal{M}_{13}+\cdots+f_n \mathcal{M}_{1n}}{f_1} & \mathcal{M}_{12} & \cdots & \mathcal{M}_{1n} \\
 \mathcal{M}_{12} & -\frac{f_1 \mathcal{M}_{12}+f_3 \mathcal{M}_{23}+\cdots+f_n \mathcal{M}_{2n}}{f_2} & \cdots & \mathcal{M}_{2n} \\
\vdots & \vdots & \ddots & \vdots \\
\mathcal{M}_{1n} & \mathcal{M}_{2n} & \cdots & -\frac{f_1 \mathcal{M}_{1n}+f_2 \mathcal{M}_{2n}+\cdots+f_{n-1} \mathcal{M}_{n-1\,n}}{f_n}
\end{array}
\right)}~.
\label{Mnn}
\eeq
The remaining $(n-1)$ eigenvalues correspond to the non-vanishing masses of the pseudo-goldstinos $\tilde{G}^{(a)}$,  $M_{G^{(a)}}$. In the simple case where $f_1{\gg} f_2{\gg} \cdots {\gg} f_n$, the contribution to the vacuum energy is $f^2 =  \sum_{i=1}^n f^2_i\approx f_1^2$ and the massless goldstino mode is aligned with $\tilde{\eta}_1$. In this case the PGLDs $\tilde{G}^{(1)},\tilde{G}^{(2)},\cdots,\tilde{G}^{(n-1)}$ will be aligned with $\tilde{\eta}_2,\tilde{\eta}_3,\cdots,\tilde{\eta}_n$, respectively, with masses approximately given by $M_{G^{(1)}}{\approx} (f/f_2)\mathcal{M}_{12}, \cdots , M_{G^{(n-1)}}{\approx} (f/f_n)\mathcal{M}_{1n}$. In the case where  the parameters $\mathcal{M}_{ij}$ ($i{<}j$) are of comparable size, as a consequence of the hierarchy $f_2{\gg} f_3{\gg} \cdots {\gg} f_n$, the masses of the PGLDs will be hierarchically ordered according to $M_{G^{(1)}}{\ll} M_{G^{(2)}}{\ll} \cdots {\ll} M_{G^{(n-1)}}$. In other words, the heaviest PGLD will be the one that is aligned with the $\tilde{\eta}_i$ arising from the hidden sector with the smallest SUSY breaking scale $f_i$. Note that the PGLD masses cannot be chosen arbitrarily large since that would imply a too large backreaction of the visible sector on the hidden sector with the smallest SUSY breaking scale~\cite{Argurio:2011hs}. In this paper we will consider the PGLD masses to be in the range \mbox{$1-300$ GeV}, and always smaller than the mass of the LOSP.

Let us now turn to the coupling of the relevant fields. We will be
interested in the case where the LOSP is a Bino-like
neutralino. Therefore we focus our attention on the couplings between
the Bino and the PGLDs. Since the mixing between the MSSM neutralinos
and $\tilde{\eta}_i$ in~\eqref{offdiag} are small, we write the rotation
matrix approximately as
\beq
      \tilde{\eta}_i \approx \frac{f_i}{f} \tilde{G} + \sum_{a=1}^{n-1} V_{ia} \tilde{G}^{(a)} \label{roteta}
\eeq
for some $n\times (n-1)$ matrix $V_{ia}$.
In (\ref{roteta}) there are also extra terms involving the neutralinos, but they are suppressed by powers of $1/f_i$ and play no role in the following since they give rise to interactions that are subleading with respect to those already present.
On the contrary, we must retain the dependence of the Bino on the PGLDs, since these provide the leading interactions that mediate the decays between the PGLDs.
Under the assumption that the off-diagonal terms
are small and that the Wino and the Higgsino masses are large, the rotation can be approximated as
\begin{equation}
   \tilde B \approx \tilde \chi + \frac{\langle D_Y \rangle}{\sqrt{2} f} \tilde G + \sum_{a=1}^{n-1} U_{a}\tilde G^{(a)}~,
   \label{goldshift}
\end{equation}
where the coefficients $U_a$ can be determined by diagonalizing (\ref{M}).
Note that they are proportional to the hypercharge $D_Y$-term VEV, and they depend on the
ratio of the contribution of each hidden sector to the Bino mass over the SUSY breaking scale.

In order to obtain the relevant couplings of the mass eigenstate neutralino $\tilde \chi$
and PGLD's $\tilde G^{(a)}$, we then take the last term of
\eqref{Bino}, sum over all
the SUSY breaking sectors, and use~\eqref{roteta} and~\eqref{goldshift}.
The couplings involving the neutralino $\tilde \chi$ are
\footnote{Note that there is also a contribution to the coupling among the lightest neutralino, a photon and a PGLD/GLD arising from the term analogous to the last term in \eqref{Bino}, but proportional to the Wino mass. However, since we assume the Winos are effectively decoupled, the neutral Wino component in the lightest neutralino will be very small and hence the contribution to the term in \eqref{Bint} is sub-dominant.}
\begin{equation}
\label{Bint}
\frac{i\,\cos \theta_w}{2\sqrt{2}}\frac{M_{B}}{f} \,\tilde{\chi}\, \sigma^{\mu}\bar{\sigma}^{\nu} F_{\mu\nu}
\left(  \tilde{G}+
\sum_{i=1}^n \sum_{a=1}^{n-1} \frac{M_{B(i)}}{M_{B}}\frac{f}{f_i} V_{ia}\, \tilde{G}^{(a)}  \right)~.
\end{equation}
In this way we obtain the usual GLD coupling as well as the additional couplings to the PGLD that differ only by an overall coefficient. Due to the presence of the ratios $M_{B(i)}/M_{B}$ and $f/f_i$ in the couplings of the PGLDs, it is possible for the PGLD couplings to be enhanced with respect to the GLD coupling. For example, in the case of direct gauge mediation, the Bino soft mass scales like $\alpha \sqrt{f_i}$.\footnote{Instead, in minimal gauge mediation, the Bino soft mass scales like $\frac{\alpha}{4\pi}\frac{f_i}{M_i}$, where $M_i$ denotes the mass of the messengers in the $i$:th hidden sector.} In this case, the PGLD couplings in \eqref{Bint} scale as $\sqrt{f/f_i}$ and hence the PGLD with the largest coupling is the one that is aligned with the $\tilde{\eta}_i$ associated with the smallest $f_i$.

Similarly, from the same term in \eqref{Bino}, we can extract the couplings among the mass eigenstate PGLD's as
\begin{equation}
   \frac{i\,\cos \theta_w}{2\sqrt{2}}\frac{M_B}{f} \left( \sum_{a=1}^{n-1} U_a\tilde G^{(a)} \right) \, \sigma^{\mu}\bar{\sigma}^{\nu} F_{\mu\nu}\left(  \tilde{G}+
\sum_{i=1}^n \sum_{b=1}^{n-1} \frac{M_{B(i)}}{M_{B}}\frac{f}{f_i} V_{ib}\, \tilde{G}^{(b)}  \right)~.
   \label{gogo}
\end{equation}
Notice that the coupling involving the same PGLD cancels because the Lorentz structure $F_{\mu\nu}(\sigma^\mu\bar\sigma^\nu)_{\alpha\beta}$ is symmetric and this is why such terms are not present in the case of a single PGLD.

The explicit expressions for $V_{ia}$ and $U_a$ are not relevant for our analysis since we will take a phenomenological approach and treat the overall coefficient of the interaction term as a free parameter.

\section{Simplified models of GMSB with goldstini}
\label{simp}

In the remainder of this paper we consider two
simplified models of GMSB with a Bino-like neutralino LOSP
$\tilde{\chi}$.
In this section we define and give motivations for these simplified models.

\begin{figure}
\center
 \includegraphics[width=.35\textwidth]{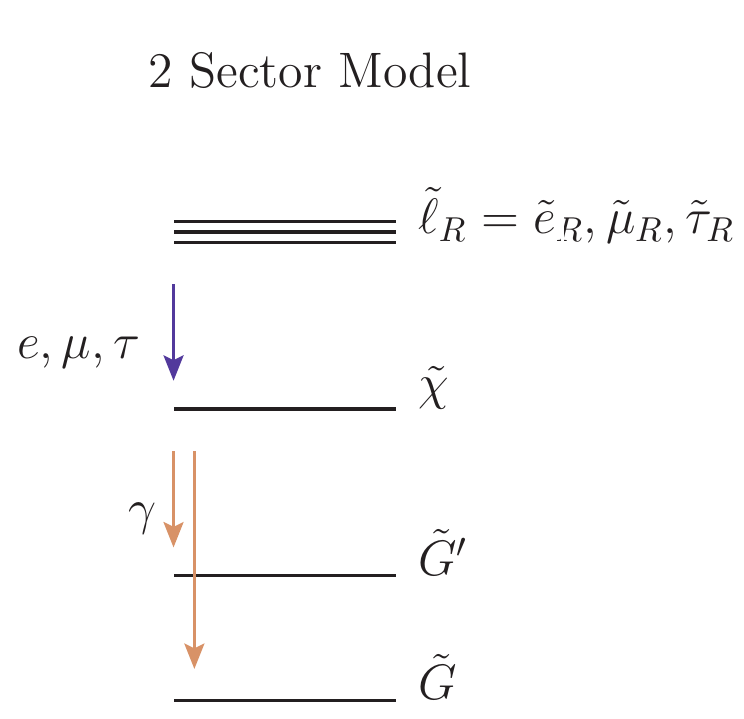}
 \qquad
 \includegraphics[width=.35\textwidth]{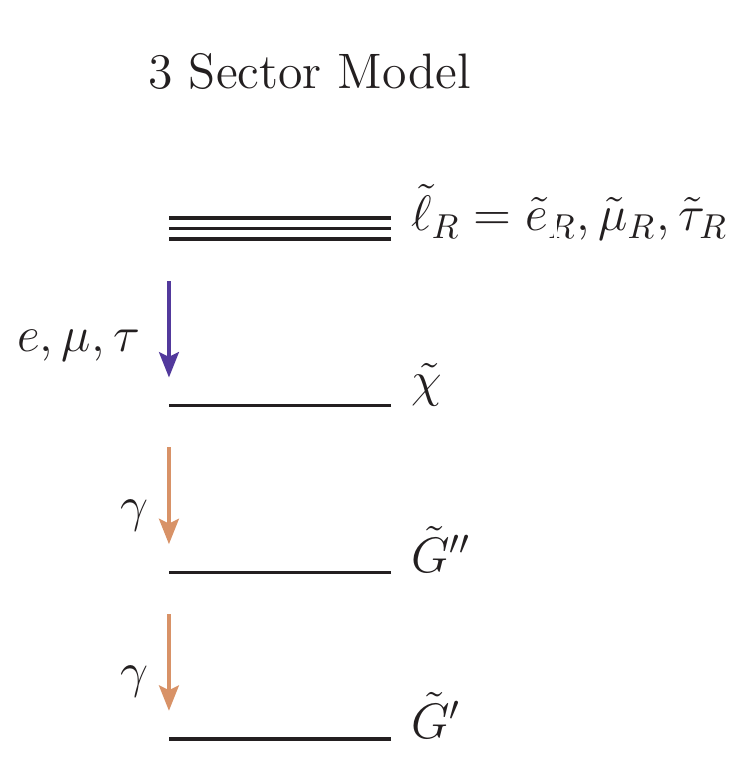}
 \caption{\small Mass spectra for the two simplified models under
 consideration.}
\label{fig:spectra}
\end{figure}

The first model consists of two hidden sectors,
i.e.~with only one PGLD $\tilde{G}'$, to which the neutralino LOSP dominantly decays, $\tilde{\chi}\to\gamma\tilde{G}'$. The second model comprises three hidden
sectors and thus two PGLDs, denoted by $\tilde{G}''$ and $\tilde{G}'$,
with masses $M_{G''}>M_{G'}$.
In this case 
the LOSP dominantly decays to a photon and the heaviest PGLD,
$\tilde{\chi}\to\gamma\tilde{G}''$, which, in turn, dominantly decays to
another photon and the lighter PGLD, $\tilde{G}''\to\gamma\tilde{G}'$.
A
schematic structure of the spectra and decay modes for the two simplified models
is given in
Figure~\ref{fig:spectra}. 
Note that in the 2 Sector Model we have shown in the spectrum also the GLD $\tilde G$,
which is relevant for the decay process of the neutralino in the case in which the mass of the PGLD $\tilde G'$
is close to the neutralino mass, as we will discuss later.
In the 3 Sector Model, instead, we will not consider benchmarks with squeezed spectra and hence the GLD $\tilde G$
is effectively irrelevant for the collider phenomenology, and not shown in Figure~\ref{fig:spectra}.

In GMSB, as a consequence of both the boundary values of the soft masses and the renormalisation group (RG) evolution, the colored superpartners are generically significantly heavier than the uncolored ones. Moreover, in order to a accommodate a Higgs mass of 125 GeV, the stops and gluinos are typically required to be in the multi-TeV range.  Among the electroweak superpartners, the right-handed sleptons and the Bino are the only ones not charged under $SU(2)_L$, and therefore they are generically the lightest SM superpartners. This motivates us to consider a simplified model where the only SM superpartners are the right-handed sleptons and the Bino-like neutralino LOSP.

The three generations of right-handed sleptons carry the same gauge
quantum numbers, implying that they are all mass degenerate at the
messenger scale. Since the Yukawa couplings enter in the RG equations,
as well as in the off-diagonal elements in the corresponding mass
matrices, it is expected that only the selectron and the smuon remain
approximately mass degenerate at low energies, whereas the lightest stau
mass eigenstate is lighter. However, unless $\tan\beta$ (and the
left-right stau mixing) is large, this splitting is small. Moreover, in
the simplified model we consider, since all three slepton generations
will dominantly decay to the Bino-like neutralino LOSP and their
corresponding SM partner, apart from a possible slight difference
in the production cross section, this small mass splitting does not
modify the phenomenology. Therefore, we take all three slepton
generations to have a common mass $M_{\ell_R}>M_{\chi}$, and the common
notation $\tilde{\ell}_R=\tilde{e}_R$, $\tilde{\mu}_R$ or
$\tilde{\tau}_R$, as shown in Figure~\ref{fig:spectra}.

We now discuss each decay step. Since the couplings of the sleptons to
the PGLDs and the GLD are strongly suppressed with respect to their
(gauge) couplings to the neutralino, the branching ratio for the
sleptons to their corresponding SM partner and the neutralino is
100\%. Hence, the first step of the chain is the same for the two-sector
and the three-sector model.

\subsection{The decay of the neutralino}\label{sec:ndecay}

\begin{figure}
\center
 \includegraphics[width=.4\textwidth]{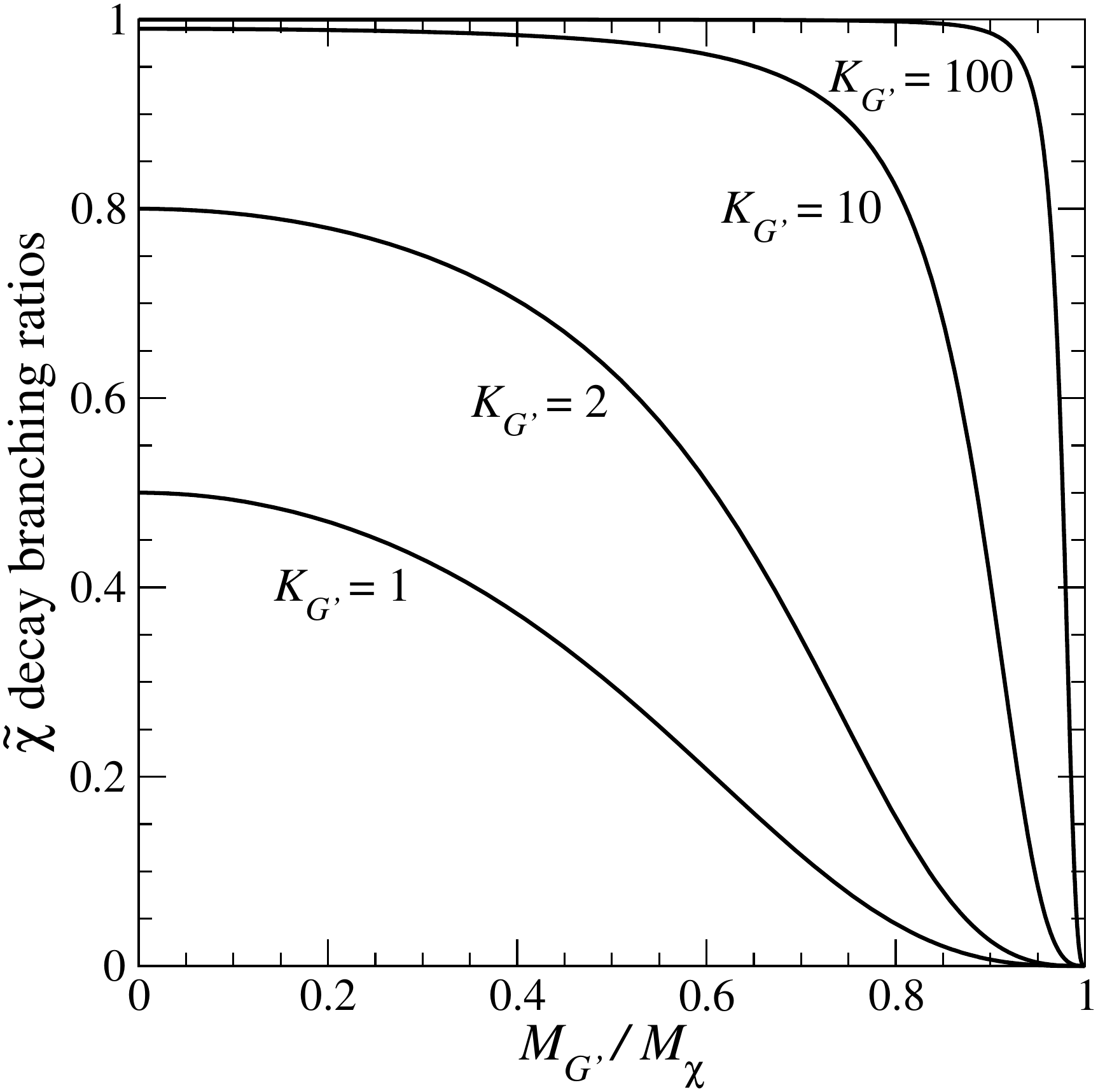}
 \caption{\small Branching ratios of $\tilde\chi\to\gamma\tilde G'$
 as a function of the mass ratio between the pseudo-goldstino and the
 neutralino for various $K_{G'}$ values. \label{fig:binodecay}}
\end{figure}

The structure of the interaction between a neutralino LOSP and a
PGLD/GLD for the two sector model was presented
in~\cite{Argurio:2011hs,Argurio:2011gu}. As was discussed in the previous
section, since the couplings and masses
of the PGLDs are strongly model-dependent, we treat them as free
parameters. We write the relevant part of the effective Lagrangian in the following way,
\beq
{\cal L}_{2} \supset
\frac{i\,\cos \theta_w}{2\sqrt{2}}\frac{M_{\chi}}{f} \,\tilde{\chi}\, \sigma^{\mu}\bar{\sigma}^{\nu} F_{\mu\nu}
\left(  \tilde{G}+K_{G'}\tilde{G}'  \right)+ \mathrm{h.c.}~,
\label{chiphoton2sect}
\eeq
leading to the partial decay width
\beq
\label{neud2sect}
     \Gamma({\tilde{\chi}}\to \gamma\,\tilde{G}') = \frac{K_{G'}^2
     \cos^2\theta_w M_{\chi}^5}{16\pi f^2} \left(1 -
     \frac{M_{G'}^2}{M_{\chi}^2} \right)^3~.
\eeq
In Figure~\ref{fig:binodecay} we show the branching ratios for the
neutralino LOSP decaying into a photon and $\tilde G'$. 
Since the
neutralino decay to the PGLD is enhanced by the factor
$K_{G'}^2$ with respect to standard GMSB, the neutralino decay will
always be prompt in the range of parameters we
are interested in, namely for $\sqrt{f}<100$ TeV. For large $K_{G'}$
the neutralino decays exclusively to the PGLD, except in the region where
the mass difference is very small. As we will see later, the two
parameters $M_{G'}$ and $K_{G'}$ give rise to richer structure in the
final state with respect to the standard GMSB signature.

Note that the neutralino decay ${\tilde{\chi}}\to Z\tilde{G}'$ is suppressed
both by phase space and by the factor
$\sin^2\theta_w/\cos^2\theta_w$. We do not consider this decay channel
in the remainder of this paper, see~\cite{Argurio:2011gu} for more details about the $Z$ decay mode.

In the three sector model, assuming $f_1 {>} f_2 {>} f_3$, the GLD will be mostly aligned with the $\tilde{\eta}_1$. Since $\eta_3$ is associated with the smallest SUSY breaking scale $f_3$, the neutralino will couple most strongly to  $\tilde{\eta}_3\approx \tilde{G}''$. The couplings and the leading decay widths are just a straightforward generalization of the two sector model discussed above, namely
\beq
{\cal L}_{3} \supset
\frac{i\,\cos \theta_w}{2\sqrt{2}}\frac{M_{\chi}}{f} \,\tilde{\chi}\, \sigma^{\mu}\bar{\sigma}^{\nu} F_{\mu\nu}
\left(  \tilde{G}+K_{G'}\tilde{G}' + K_{G''}\tilde{G}''  \right) + \mathrm{h.c.}\,,
\label{chiphoton}
\eeq
where $K_{G''}>K_{G'}>1$, with the decay width formula for ${\tilde{\chi}}\to \gamma\,\tilde{G}''$ being the same as in~\eqref{neud2sect}, upon the replacements $K_{G'}\to K_{G''}$ and $M_{G'}\to M_{G''}$. The neutralino decay to the GLD, is also given by \eqref{neud2sect}, upon the replacements $K_{G'}\to 1$ and $M_{G'}\to 0$.

\subsection{The decay of the pseudo-goldstino}

We now turn to the third step of the cascade, i.e.~to the $\tilde{G}''$ decay
that is relevant for the three sector model in Figure~\ref{fig:spectra}. The leading decay channel is $\tilde{G}''\to\gamma\tilde{G}'$, arising via the last operator in \eqref{Bino}, through the Bino component of one of the two PGLDs. This mixing between the Bino and the PGLDs arises from the $\tilde{B}$-$\tilde{\eta}$ mixing term in~\eqref{Bino}, which gives rise to the off-diagonal neutralino mass matrix in~\eqref{M}, i.e.~the first row in \eqref{offdiag}. 
Since we will only consider the case where the mass splitting between the PGLDs is less than 100~GeV, the decay $\tilde{G}''\to Z\tilde{G}' $ is strongly phase space suppressed (if open at all) and we do not consider it in this paper.


The decay width $\Gamma(\tilde{G}''\to\gamma\tilde{G}')$ depends on many parameters, including the MSSM parameters in \eqref{M44} and \eqref{offdiag}, as well as those in \eqref{Mnn}. In order to give an ``existence proof" that the decay $\tilde{G}''\to\gamma\tilde{G}'$ can be prompt, we provide explicit examples of parameter choices that give rise to prompt decays. Our parametrization is inspired by direct gauge mediation, as we set the contributions to the Bino and Wino masses from the $i$:th hidden sector to be $M_{B(i)}=c_B \alpha_1\sqrt{f_i}$ and $M_{W(i)}=c_W \alpha_2 \sqrt{f_i}$, where $i=1,2,3$. We have set $c_B=1/2$ and since we are interested in a simplified model where the Winos are effectively decoupled, we have taken $c_W$ to be one order of magnitude larger than $c_B$. The soft Higgs down/up type masses are set to be $m_{H_{d/u}(i)}^2=(f_i/f)m_{H_{d/u}}^2$, where $f^2=\sum_{i=1}^3f_i^2$ and $m_{H_{d/u}}^2$ are the total soft Higgs down/up type masses, obeying the usual MSSM electroweak symmetry breaking conditions, for which we have used $B_{(i)}=(f_i/f)B$ with $B=(500 \textrm{\,GeV})^2$, $\mu=1$ TeV and $\tan\beta=10$ as input parameters. 

The SUSY breaking scales are fixed to be $\sqrt{f_1}=28$ TeV, $\sqrt{f_2}=2.5$ TeV and $\sqrt{f_3}=0.5$ TeV.
The hierarchy between these scales is necessary in order to obtain PGLD masses in the range we will consider in this paper.  These values imply a Bino mass of around 150 GeV, which is a typical value we will use in the benchmark points analyzed in Section \ref{3sectorsec}.

Our choice of parameters determines the full 7$\times$7 neutralino mass matrix in \eqref{M} up to the 
entries associated to the
PGLD's \eqref{Mnn}, which are characterized by the three unknown two point functions $\mathcal{M}_{12}$, $\mathcal{M}_{13}$ and $\mathcal{M}_{23}$.
We diagonalize the neutralino mass matrix and 
parameterize these entries in terms of the 
three smallest eigenvalues. They correspond to the mass eigenvalues for the $\tilde G''$, $\tilde G'$
and $\tilde G$, respectively, where the last eigenvalue is zero. 
We then scan over the $\tilde G''$ and $\tilde G'$ masses in the range $m_{G'}=\{0,50\}$ GeV and $m_{G''}=\{m_{G'}+20,100 \}$ GeV. We have checked that each of these points can be mapped to values for the two point functions $\mathcal{M}_{ij}$ which are consistent with 
the perturbative computation in \cite{Argurio:2011hs}.

\begin{figure}
 \center
 \includegraphics[width=.6\textwidth]{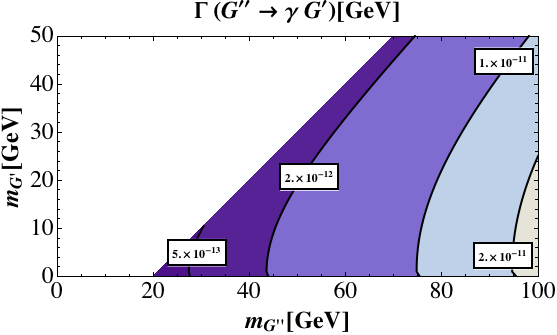}
 \caption{\small The partial decay width 
　$\Gamma(\tilde{G}''\to\gamma\tilde{G}')$ as a function of the $\tilde{G}''$ and $\tilde{G}'$ masses. \label{G3G2}}
\end{figure} 

In Figure \ref{G3G2} we show the result of the numerical scan in terms of the width $\Gamma(\tilde{G}''\to\gamma\tilde{G}')$, as a function of the $\tilde{G}''$ and $\tilde{G}'$ masses, obtained from the couplings to the photon in the mass eigenbasis. As can be seen from this figure, for our choice of parameters, there is a large region for which the decay width is greater than $2\times 10^{-12}$ GeV, i.e.~for which $c\tau$ is smaller than \mbox{0.1 mm}. At the same time, there is no point in this mass plane for which the decay of $\tilde G'$ to $\tilde G$ occurs inside the detector.  
Moreover, in this mass plane,
the branching ratio BR$({\tilde{G}''}\to \gamma \tilde{G}') $ is always close to $100 \%$, while the 
BR$({\tilde{\chi}}\to \gamma \tilde{G}'') $ varies from around 85$\%$, for large value of $m_{G''}$, to nearly 100$\%$, for small $m_{G''}$.

Note that $\tilde{G}''$ could also have three-body decays such as
$\tilde{G}''\to e^+e^-\tilde{G}'$ (or any other fermion pair) and
$\tilde{G}''\to\gamma\gamma\tilde{G}'$ (or any other vector-boson
pair). However we compute these decays analytically in the Appendix and find that none of
them can be prompt. (Even though these three-body decays are not relevant for
collider physics, they can be useful in the context of cosmology for
gravity mediation scenarios~\cite{Cheng:2010mw}.)

\section{Signatures at the LHC}\label{sign}

\begin{figure}
 \center
 \includegraphics[width=.5\textwidth]{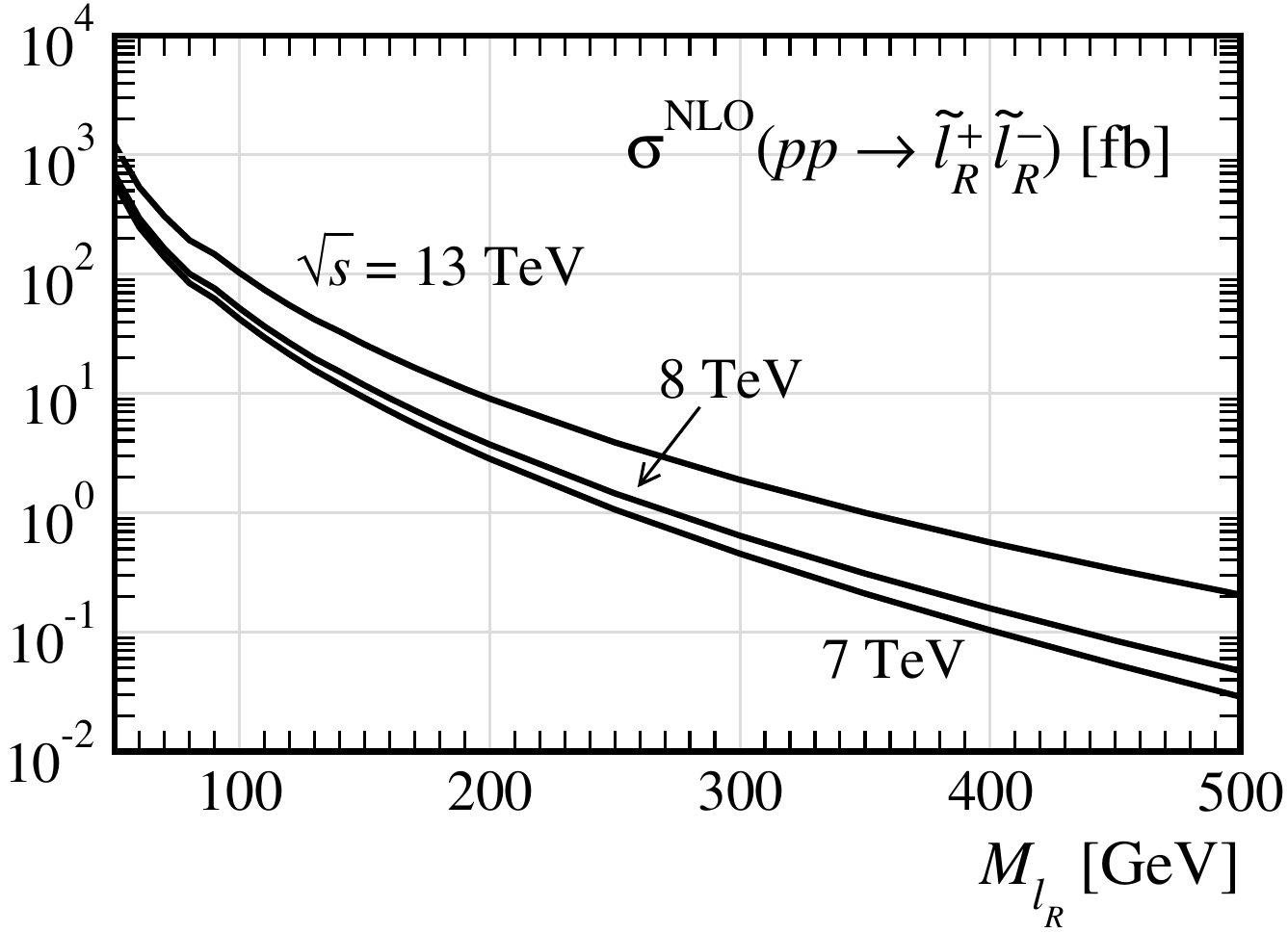}
 \caption{\small Right-handed slepton pair-production cross section
 at the LHC, for a single flavor, as a function of the slepton
 mass. \label{fig:xsec}}
\end{figure} 

In this section we discuss the signatures and phenomenology of the two
simplified GMSB models described in the previous section
with their respective mass spectrum given in
Figure~\ref{fig:spectra}. In both of these models, the relevant
production mode is slepton pair production,
$pp\to\tilde{\ell}_R^{+}\tilde{\ell}_R^{-}$, via the electroweak
Drell-Yan process. The cross sections at the LHC for $\sqrt{s}=7$, 8 and
13~TeV are provided in Figure~\ref{fig:xsec}, as computed by
{\sc MadGolem}~\cite{Binoth:2011xi} at the next-to-leading order in QCD. 
In the two sector model, the slepton pair production process gives rise
to the final state $\ell^+\ell^- +2\gamma+\MET$, see
Figure~\ref{fig:stableproc}, whereas in the three sector model, the
final state is $\ell^+\ell^- +4\gamma+\MET$, see
Figure~\ref{fig:unstableproc}.
Recall that we are using a common notation for the sleptons,
$\tilde{\ell}_R^{\pm}=\tilde{e}_R^{\pm},\tilde{\mu}_R^{\pm},\tilde{\tau}_R^{\pm}$,
as well as for the leptons, $\ell^{\pm}=e^{\pm},\mu^{\pm},\tau^{\pm}$.

\subsection{The two sector model}

\begin{figure}
\centering
\subfigure[]{%
{ \includegraphics[width=.4\textwidth]{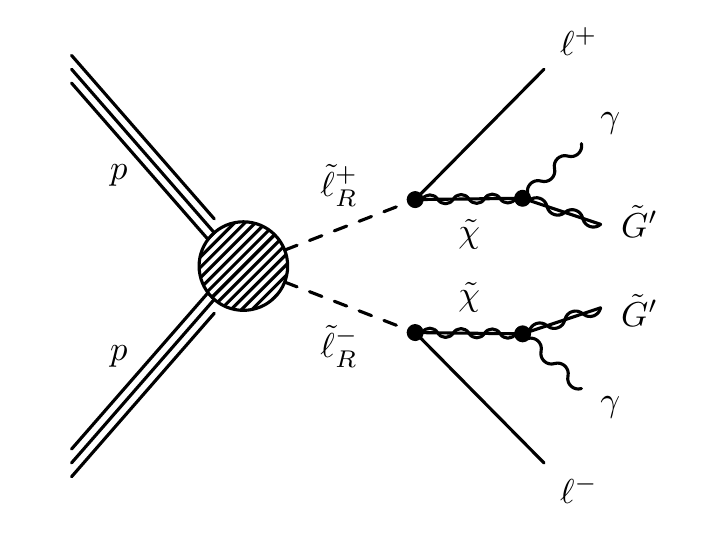}
  \label{fig:stableproc}}}
\quad
\subfigure[]{%
 \includegraphics[width=.4\textwidth]{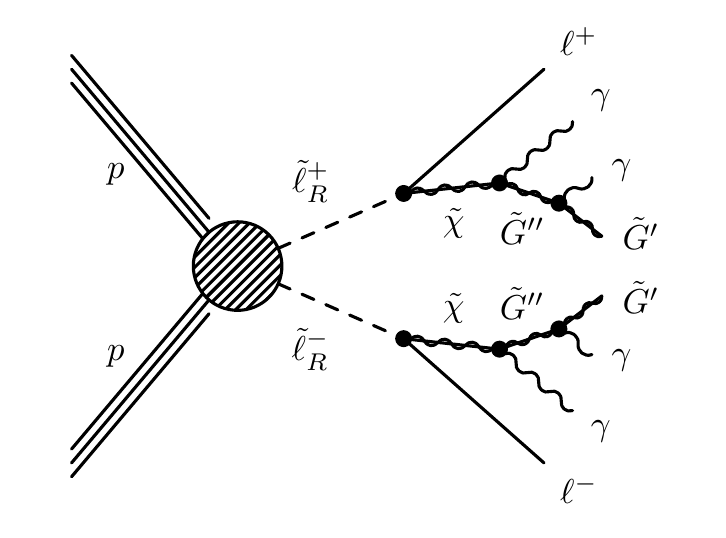}
  \label{fig:unstableproc}}
\caption{\small The processes relevant at the LHC in the (a) two- and (b)
three-sector models, corresponding to the spectra in
 \mbox{Figure~\ref{fig:spectra}}.}
\label{fig:proc}
\end{figure}

We begin our collider study by considering the two sector model. This
model has been studied for the specific SPS8 point for the LHC-14TeV
in~\cite{Argurio:2011gu}. Here, we extend the analysis to a generic
model parameter space and
confront it with currently available LHC searches. Furthermore, at the end of this section, we propose new
search strategies designed to probe this scenario.

The most relevant LHC search we have found is the inclusive
diphoton+$\MET$ search done by ATLAS~\cite{Aad:2012zza}, which we will
describe below. Also the CMS collaboration has performed searches for
diphoton+$\MET$~\cite{CMSdiphotons} but due to the jet requirements, these
searches are less sensitive to our models, where jets only arise from
initial state radiation. There is also a CMS dilepton+$\MET$
search~\cite{CMSdileptons} that in principle could be sensitive to our
model. However, due to the large SM diboson background, the efficiency
in this search drops fast as the mass of the neutralino is within around
100~GeV of the slepton mass. Since, in our models, the neutralino is
generically within this region, the search~\cite{CMSdileptons} is not
sensitive. There is also an ATLAS dilepton+$\MET$
search~\cite{ATLASdileptons} but due to a jet veto, and since photons
are counted as jets in this search, there is an implicit veto on
photons.%
\footnote{We thank Beate Heinemann, Andreas Hoecker and Monica D'Onofrio
 for helpful discussions concerning the ATLAS searches mentioned in this
 paragraph.}
Finally, there are ATLAS searches for
$\ell+\gamma+\MET$~\cite{ATLASphotonlepton} and for $\gamma+\MET$~\cite{Aad:2012fw}, but due to the tight cuts on the photon and the $\MET$, these searches are
less sensitive than the ATLAS diphoton+$\MET$ search.%
\footnote{There is also a CMS search for
 $\ell+\gamma+\MET$~\cite{Chatrchyan:2011ah}, but it is based on only 35
 pb$^{-1}$ of data at $\sqrt{s}=7$ TeV.}
Let us focus on the most relevant search, i.e.~the ATLAS
diphoton+$\MET$ search~\cite{Aad:2012zza} in the following.

For our signal simulation, we use the goldstini
model~\cite{Argurio:2011gu,Mawatari:2012ui} (building
on~\cite{Mawatari:2011jy}) implemented in
{\sc FeynRules}~\cite{Alloul:2013bka} and pass it to
{\sc MadGraph5}~\cite{Alwall:2011uj} for event generation by means of
the {\sc UFO} library~\cite{Degrande:2011ua,deAquino:2011ub}. We employ
{\sc Pythia}~\cite{Sjostrand:2006za} for parton shower and
hadronization, {\sc Delphes}~\cite{deFavereau:2013fsa} for fast detector
simulation with the ATLAS setup, and
{\sc MadAnalysis5}~\cite{Conte:2012fm} for sample analysis.

\begin{figure}
 \center
 \includegraphics[width=.78\textwidth]{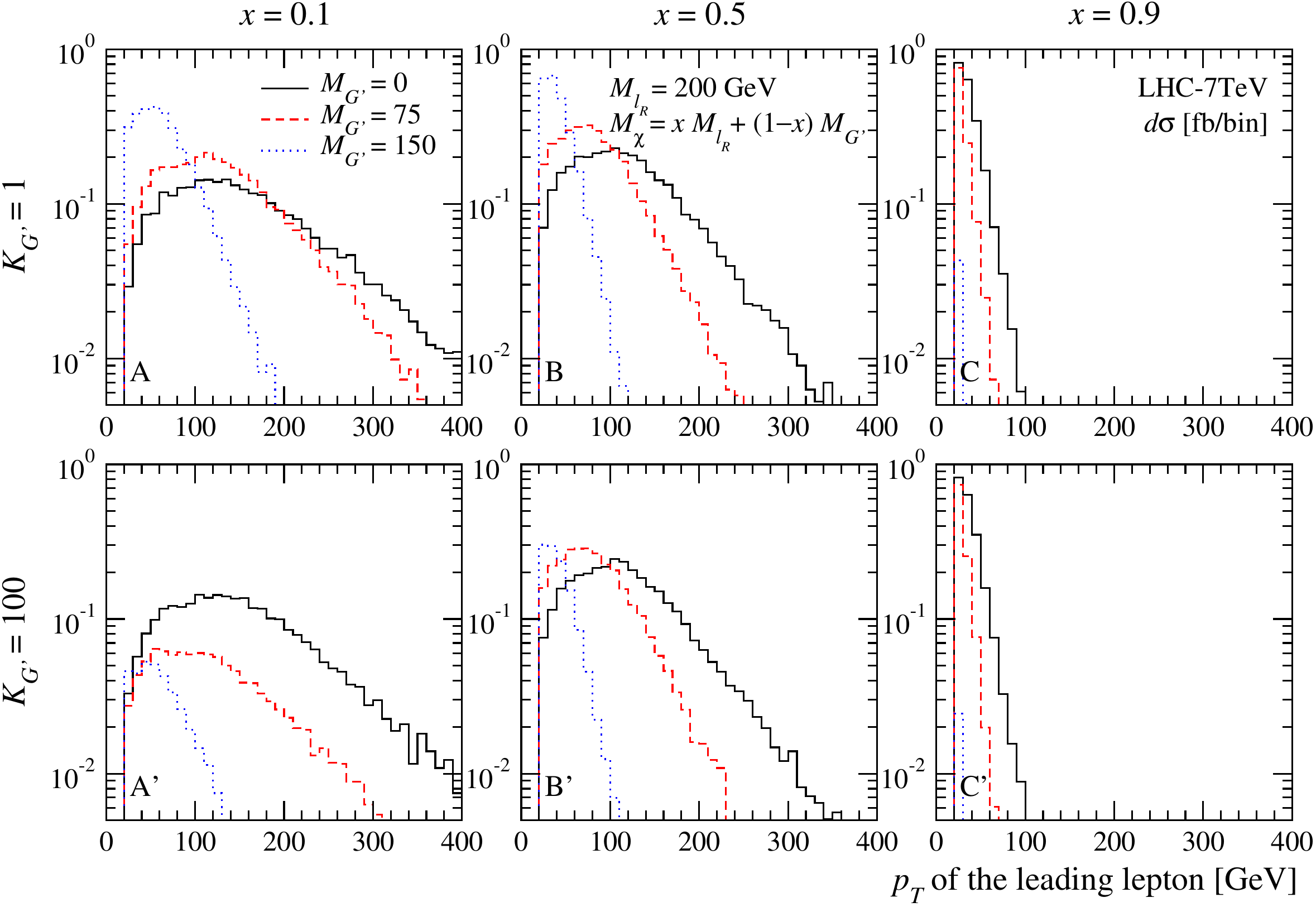}
 \caption{\small Transverse momentum distributions of the leading lepton
 for
 $pp\to\tilde\ell_R^+\tilde\ell_R^-\to\ell^+\ell^-+\gamma\gamma+\misset$
 at $\sqrt{s}=7$~TeV for $M_{\ell_R}=200$~GeV with various
 neutralino and pseudo-goldstino masses. The value of $K_{G'}$ is fixed
 at 1 and 100 in the top and bottom figures, respectively.}
\label{fig:e1}
\end{figure}

\begin{figure}
 \center
 \includegraphics[width=.78\textwidth]{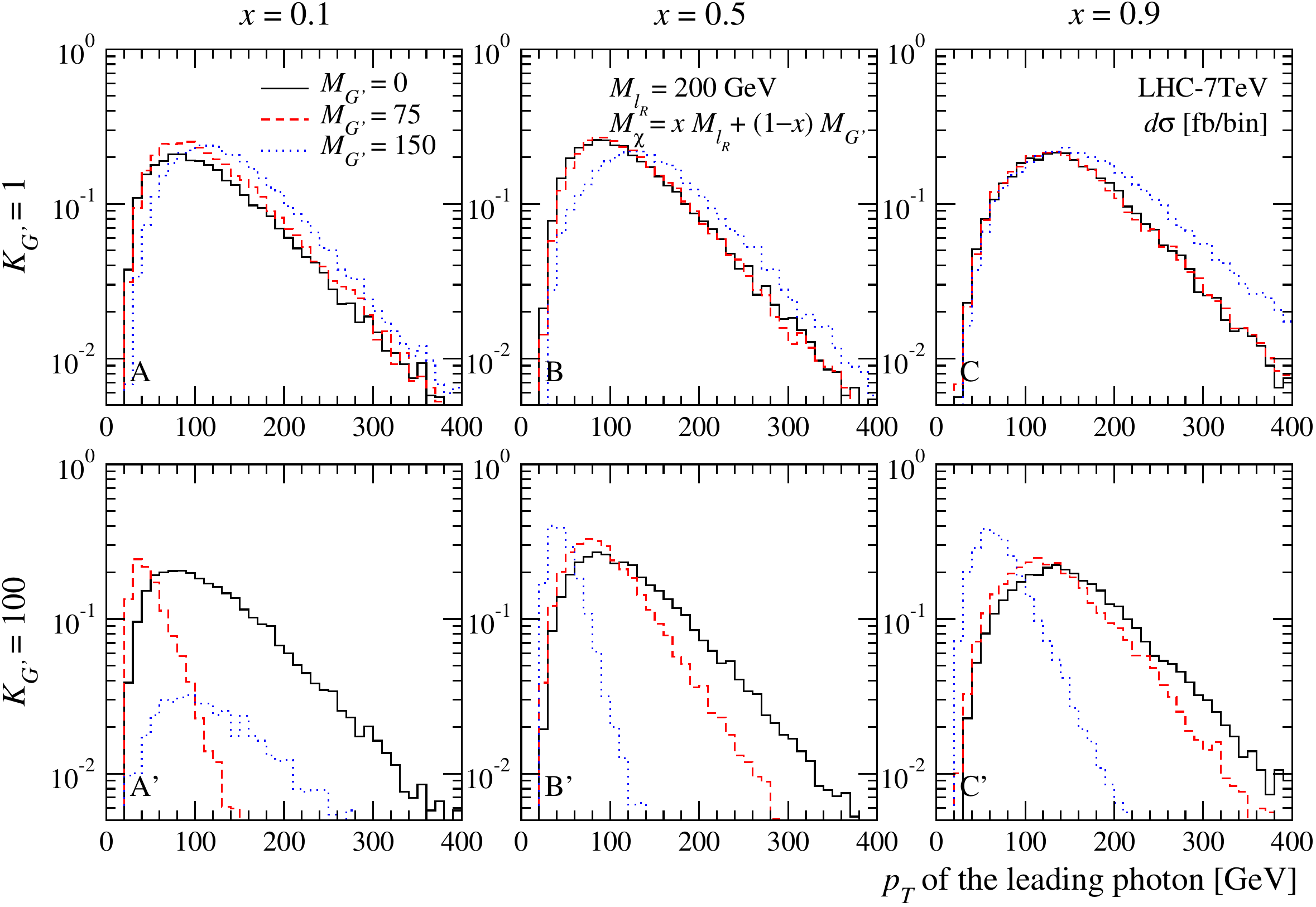}
 \caption{\small The same as in Fig.~\ref{fig:e1}, but for the
 transverse momentum of the leading photon.}
\label{fig:a1}
\end{figure}

\begin{figure}
 \center
 \includegraphics[width=.78\textwidth]{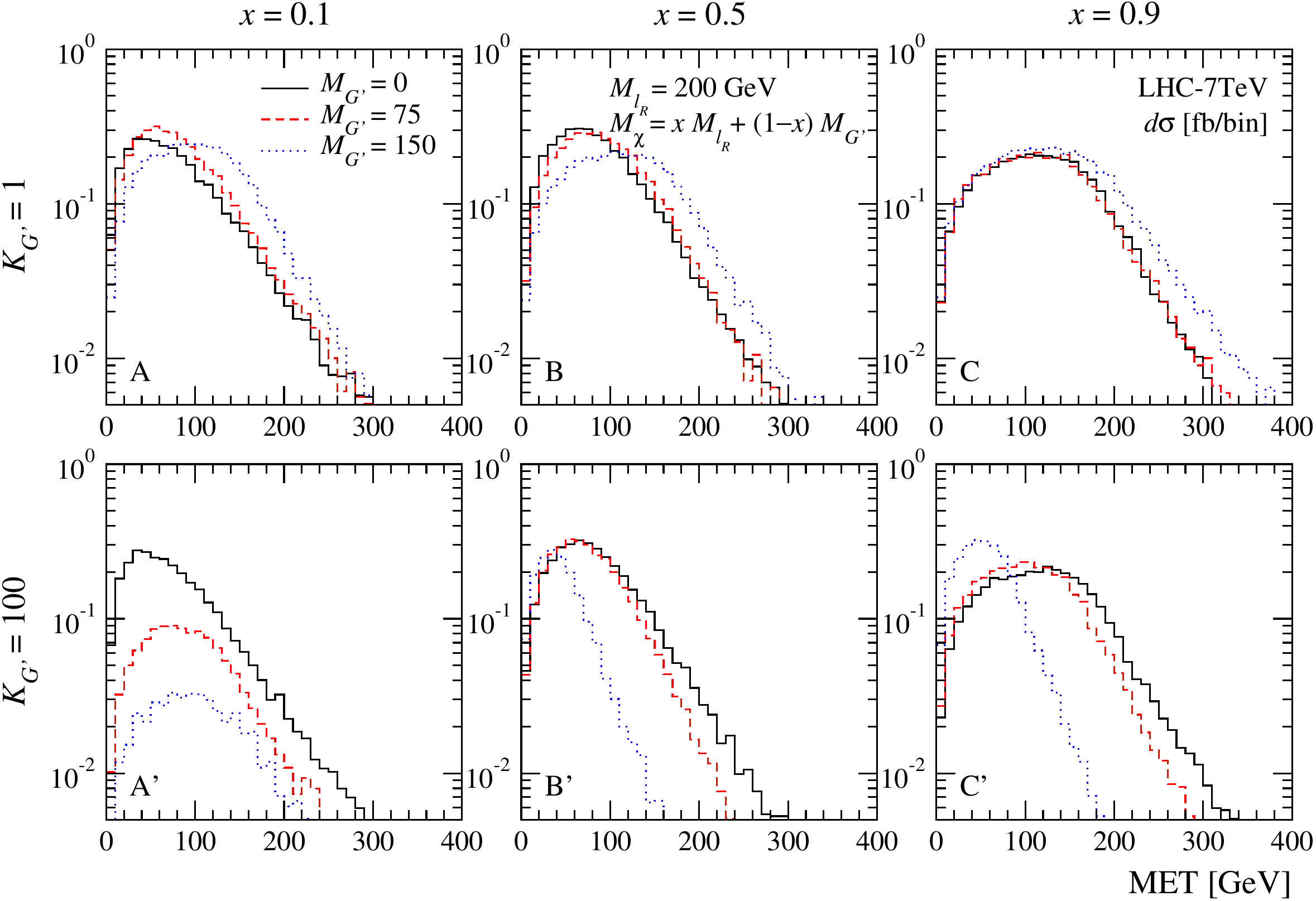}
 \caption{\small The same as in Fig.~\ref{fig:e1}, but for the missing
 transverse energy.}
\label{fig:met}
\end{figure}

\begin{figure}
 \center
 \includegraphics[width=.78\textwidth]{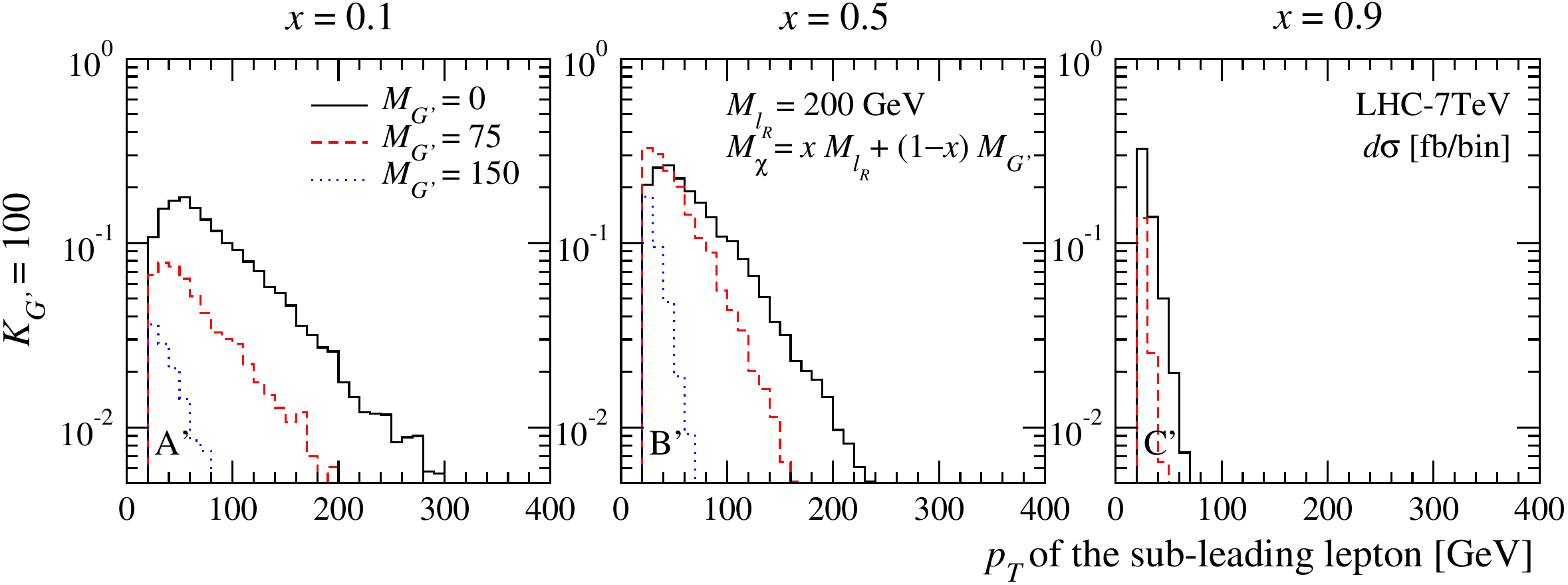}\\[2mm]
 \includegraphics[width=.78\textwidth]{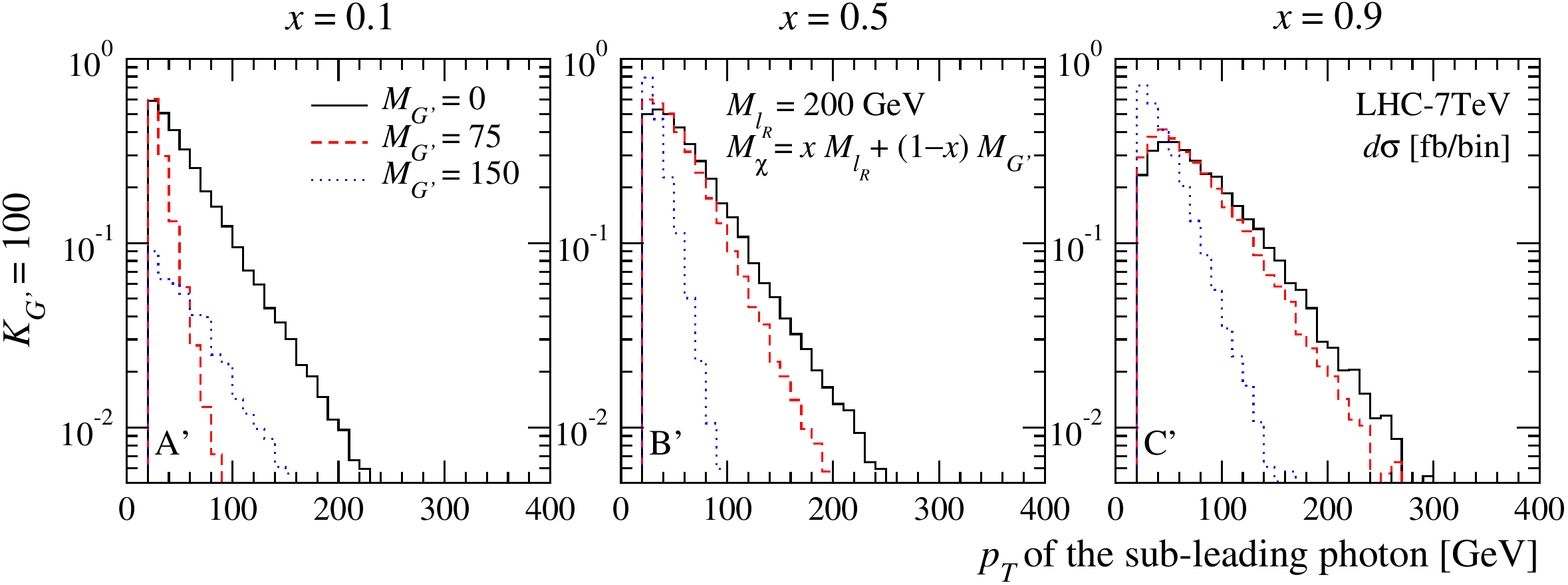}
 \caption{\small The same as in Fig.~\ref{fig:e1}, but for the
 transverse momentum of the sub-leading lepton (top) and photon (bottom)
 for $K_{G'}=100$ only.}
\label{fig:e2}
\end{figure}

Here we consider the slepton pair production and the cascade decay as
shown in Figure~\ref{fig:stableproc}:
\begin{align}
 pp\to\tilde\ell_R^+\tilde\ell_R^-\to\ell^+\ell^-\tilde\chi\tilde\chi~;
 \quad
 \tilde\chi\to\gamma\tilde G'\quad{\rm or}\quad
 \tilde\chi\to\gamma\tilde G~,
\end{align}
resulting in $\ell^{+}\ell^{-}+\gamma\gamma+\MET$.
As discussed in Sec.~\ref{sec:ndecay}, whether the neutralino LOSP decays to a
PGLD or a GLD depends on the PGLD mass and the $K_{G'}$
factor in~\eqref{chiphoton}, leading to distinctive final-state spectra.
To illustrate the parameter dependence, we present the
kinematic distributions for $\sqrt{s}=7$~TeV in
Figures~\ref{fig:e1} to \ref{fig:e2}, where we
fix the slepton mass at $M_{\ell_R}=200$~GeV and vary the PGLD mass as
$M_{G'}=0,75,150$~GeV. The neutralino mass is taken to be between the
slepton and the PGLD as $x=0.1,0.5,0.9$ with
\begin{align}
\label{nmass}
 M_{\chi}= xM_{\ell_R}+(1-x)M_{G'}~,
\end{align}
corresponding to the three cases where the neutralino mass is either
close to the PGLD mass, halfway between the PGLD and the slepton, or
close to the slepton mass, respectively. The two cases $K_{G'}=1$ and
100 are shown in Figures~\ref{fig:e1} to \ref{fig:met}, while for the
sub-leading lepton and photon spectra in Figure~\ref{fig:e2} we present
only $K_{G'}=100$. Here the isolated leptons and photons are required
to pass the following minimal detector cuts:
\begin{align}
 &p_T^{\ell}>20~{\rm GeV}~,\quad |\eta^{\ell}|<2.5~, \label{lcut}\\
 &p_T^{\gamma}>20~{\rm GeV}~,\quad |\eta^{\gamma}|<1.81
 \quad({\rm except}\ 1.37<|\eta^{\gamma}|<1.52)~, \label{acut}
\end{align}
where the transition region between the barrel and end-cap calorimeters
is taken into account for photons according to the ATLAS
search~\cite{Aad:2012zza}. We also require at least two photons in the
final state.

The different $p_T$ distributions of the leading lepton for the
different benchmark points, shown in Figure~\ref{fig:e1}, depend on
the mass difference in the first cascade decay, i.e. between the slepton
and the neutralino. The heavier neutralino (i.e. large $x$ and $M_{G'}$)
the softer the lepton becomes.
While the shape of the distributions does not depend on the $K_{G'}$
value, the normalization does due to the efficiency of the minimal
detector cuts in~\eqref{lcut} and \eqref{acut}.
We note that the $M_{G'}=0$ (nearly massless) case is corresponding to
the standard GMSB scenario, for which we have essentially two
indistinguishable copies of a light GLD.

On the other hand, the $p_T$ distributions for the most energetic
photon, shown in Figure~\ref{fig:a1}, depends on the mass difference in
the second decay, i.e.~between the neutralino LOSP and the PGLD/GLD.
For $K_{G'}=1$ the LOSP decay into a PGLD is phase space suppressed when
the PGLD is heavy. For instance, for $M_{G'}=150$~GeV the branching
ratio is about 10\%; see Figure~\ref{fig:binodecay}. Therefore, the
spectrum becomes harder. In contrast, for $K_{G'}=100$, the decay to the
PGLD is dominant even for PGLD masses close to the neutralino, making
the photons significantly softer. It is clear from the $p_T$
distributions of the photons in Figures~\ref{fig:a1} and
\ref{fig:e2} that in such case
a severe $p_T$ cut, e.g.~$p_T>50$~GeV on both
photons as in the ATLAS analysis~\cite{Aad:2012zza}, has a dramatic effect on the efficiency. In
order for the photons to pass such a high $p_T$ cut it is necessary for
the mass splitting between the neutralino and the PGLD to be large, such
that the emitted photons are sufficiently energetic. Moreover, since the
phase space of the invisible PGLDs is reduced for a massive PGLD in
comparison to the massless case, the amount of the missing transverse
energy is also reduced, as can be seen from the $\MET$ distributions in
Figure~\ref{fig:met}.
This reduces the efficiency of the $\MET$ selection cuts as well.

\begin{table}[t]
 \center
 \footnotesize
 \begin{tabular}{|rr||rrrr||rr|}
 \hline
  \multicolumn{2}{|r||}{$M_{\ell_R}=200$~GeV}
 & \multicolumn{4}{c||}{$2\gamma+\misset$}
 & \multicolumn{2}{c|}{$2\ell+2\gamma+\misset$} \\
  \multicolumn{2}{|r||}{($K_{G'},\,x,\,M_{G'}$)} & min
 & $+\,p_T^{\gamma_{1,2}}>50$ & $+\,\misset>50$
 & $+\,\misset>125$
 & min & $+\,\misset>50$ \\
 \hline\hline
  &(1,\,0.1,\,0)   & 0.35 & 0.17 & 0.11 & 0.03 & 0.21 & 0.14 \\
 A&(1,\,0.1,\,75)  & 0.41 & 0.22 & 0.15 & 0.04 & 0.24 & 0.17 \\
  &(1,\,0.1,\,150) & 0.42 & 0.29 & 0.24 & 0.09 & 0.17 & 0.14 \\
 \hline
  &(1,\,0.5,\,0)   & 0.42 & 0.24 & 0.18 & 0.05 & 0.24 & 0.18 \\
 B&(1,\,0.5,\,75)  & 0.42 & 0.26 & 0.20 & 0.06 & 0.21 & 0.16 \\
  &(1,\,0.5,\,150) & 0.40 & 0.29 & 0.25 & 0.12 & 0.09 & 0.08 \\
 \hline
  &(1,\,0.9,\,0)   & 0.40 & 0.30 & 0.25 & 0.13 & 0.07 & 0.06 \\
 C&(1,\,0.9,\,75)  & 0.40 & 0.29 & 0.25 & 0.12 & 0.02 & 0.02 \\
  &(1,\,0.9,\,150) & 0.50 & 0.39 & 0.34 & 0.19 & $<0.01$ & $<0.01$ \\
 \hline\hline
   &(100,\,0.1,\,0)   & 0.35 & 0.18 & 0.11 & 0.03 & 0.21 & 0.14 \\
 A'&(100,\,0.1,\,75)  & 0.14 & 0.01 & 0.01 & $<0.01$ & 0.07 & 0.06 \\
   &(100,\,0.1,\,150) & 0.06 & 0.03 & 0.03 & 0.01 & 0.01 & 0.01 \\
 \hline
   &(100,\,0.5,\,0)   & 0.42 & 0.24 & 0.17 & 0.05 & 0.24 & 0.18 \\
 B'&(100,\,0.5,\,75)  & 0.38 & 0.19 & 0.13 & 0.02 & 0.19 & 0.13 \\
   &(100,\,0.5,\,150) & 0.20 & 0.02 & 0.01 & $<0.01$ & 0.04 & 0.01 \\
 \hline
   &(100,\,0.9,\,0)   & 0.40 & 0.30 & 0.26 & 0.13 & 0.06 & 0.06 \\
 C'&(100,\,0.9,\,75)  & 0.40 & 0.26 & 0.21 & 0.08 & 0.02 & 0.02 \\
   &(100,\,0.9,\,150) & 0.30 & 0.10 & 0.05 & $<0.01$ & $<0.01$ & $<0.01$ \\
 \hline
 \end{tabular}
 \caption{\small Cumulative selection efficiencies after each requirement from
 the left to right for the different benchmark
 points of ($K_{G'},\,x,\,M_{G'}$) for the slepton mass at 200~GeV
 at the LHC-7TeV.
 In the column ``min'', the minimal cuts in~\eqref{lcut} and
 \eqref{acut} are imposed for two photons (plus two leptons) in the
 $2\gamma+\MET$ ($2\ell+2\gamma+\misset$) category.}
\label{tab:eff}
\end{table}

In order to see more explicitly how the kinematic cuts reduce the
efficiencies, we show in Table~\ref{tab:eff} the cumulative selection
efficiencies for the same benchmark points of ($K_{G'},\,x,\,M_{G'}$) as
in the distributions above. As one can see, even with the minimal
detector cuts in~\eqref{lcut} and \eqref{acut}, the efficiencies of some
benchmarks, especially for $K_{G'}=100$, where the
$\tilde\chi\to\gamma\tilde G'$ is dominant, are quite low.

Note that the efficiencies are different among the lepton flavors in
the final state, i.e. better for muons while worse for taus with
respect to electrons. Moreover, the tau decays give rise to
an additional source of $\MET$, arising from the neutrinos, but we have
checked that the difference in the $\MET$ distributions compared to the
other lepton flavors is insignificant. For simplicity we
consider only selectron pair production in our simulation.

Motivated by the ATLAS diphoton search~\cite{Aad:2012zza}, in Table \ref{tab:eff}
we impose
$p_T>50$~GeV for the leading and sub-leading photons in addition to the
minimal cuts. Moreover, an additional $\MET$ cut is imposed as
$\MET>50$ and 125~GeV. The latter $\MET$ cut is the one ATLAS imposed
for the signal region (SR) C, which is the most relevant scenario for
us. With an integrated luminosity of 4.8~fb$^{-1}$, ATLAS observed two
events for SR~C, which is in good agreement with the expected number of background
events 2.11, resulting in about five events for the 95\% CL upper limit
on the the number of events.
Since the number of the signal events for
$M_{\ell_R}=200$~GeV is expected to be about 40 times the efficiency,
the benchmark points with the efficiency of more than 0.12 are
excluded. We find that
the $p_T$ and $\MET$ cuts in the ATLAS analysis make the search poorly sensitive to
our simplified models.
The only cases which are constrained by the ATLAS search are
 the $K_{G'}=1$ case with $x=0.9$ as well as the $(K_{G'},x,M_{G'})=(100,0.9,0)$ one.
 The $x=0.9$ case is generically more promising since the $p_T$
 of the photons is larger, given the large mass difference between the
 neutralino and the PGLD (see \eqref{nmass}).
We note that the azimuthal separation cut between a
photon and the missing transverse momentum vector, imposed by ATLAS~\cite{Aad:2012zza},
reduces the efficiencies by a few percent at most, resulting in slightly
weaker exclusion bounds.

\begin{figure}[t]
 \center
 \subfigure[Efficiencies for $x=0.5$.]{%
  \includegraphics[width=.48\textwidth]{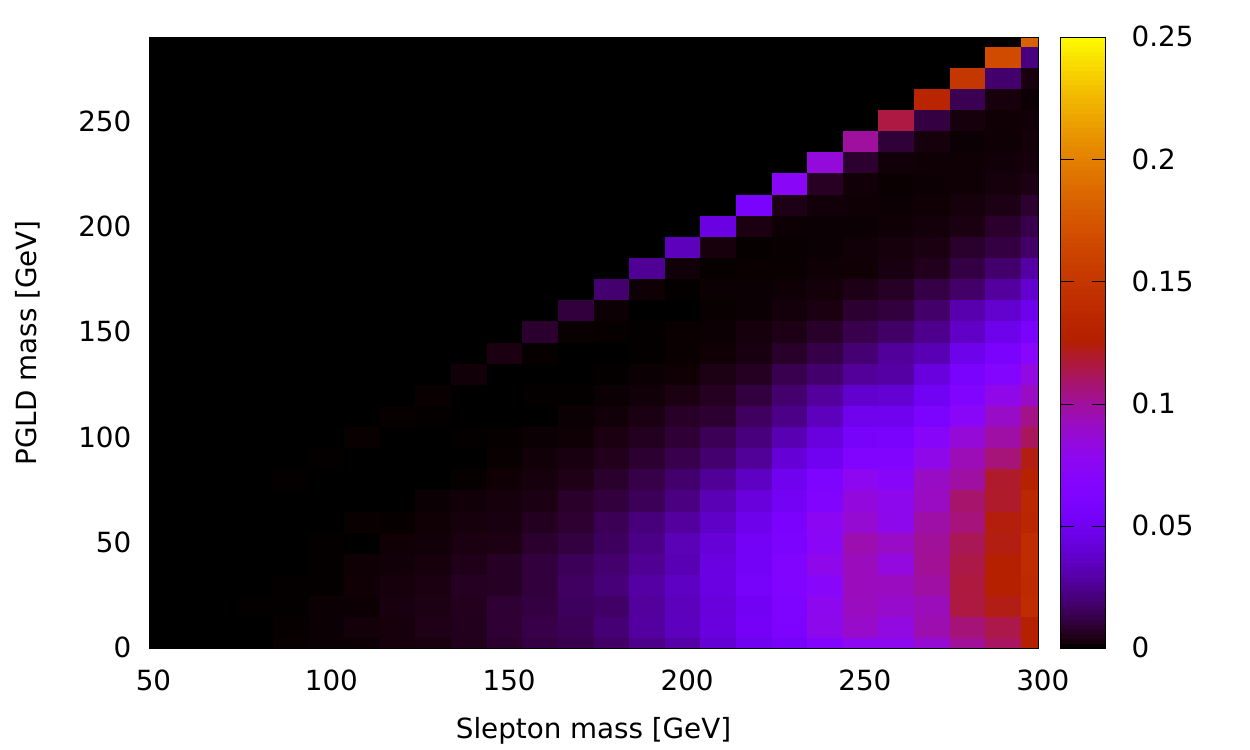}
  \label{fig:eff05}}
 \subfigure[Efficiencies for $x=0.9$.]{%
  \includegraphics[width=.48\textwidth]{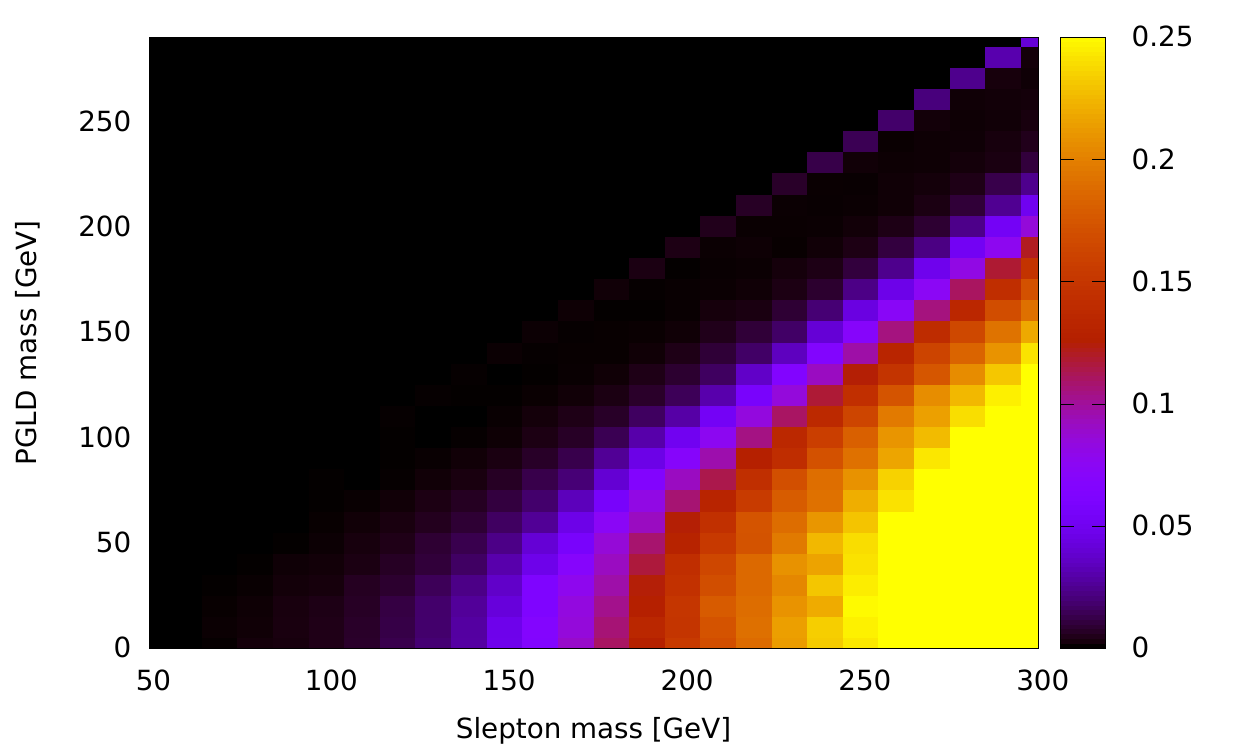}
 \label{fig:eff09}}
 \subfigure[Visible cross sections for $x=0.5$.]{%
  \includegraphics[width=.48\textwidth]{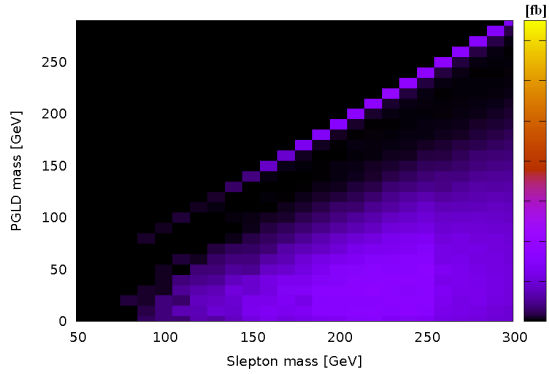}
  \label{fig:xsec05}}
 \subfigure[Visible cross sections for $x=0.9$.]{%
  \includegraphics[width=.48\textwidth]{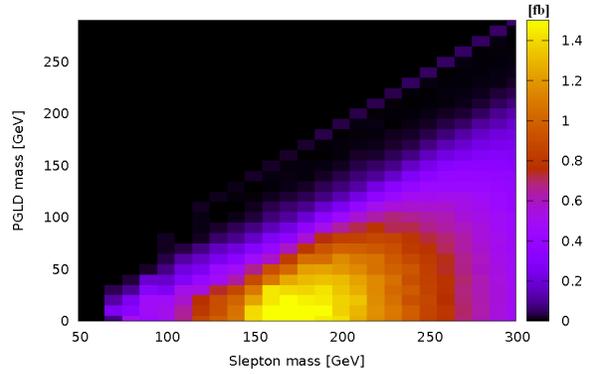}
  \label{fig:xsec09}}
 \caption{\small Selection efficiencies and visible cross
 sections in fb for
 $pp\to\tilde\ell_R^+\tilde\ell_R^-\to\ell^+\ell^-+\gamma\gamma+\misset$
 at $\sqrt{s}=7$~TeV in the ($M_{\ell_R},\,M_{G'}$) mass plane, for
 $K_{G'}=100$ with $x=0.5$ and 0.9.}
\label{fig:visxsec}
\end{figure}

To survey the entire parameter space, in Figure~\ref{fig:visxsec} we
scan over different values of the slepton mass $M_{\ell_R}$ and the PGLD
mass $M_{G'}$ for $K_{G'}=100$ with $x=0.5$ and 0.9, and present the
selection efficiencies and the visible cross sections for LHC-7TeV in
the $M_{\ell_R}$-$M_{G'}$ mass plane. For this scan we use the partonic
events and in addition to the minimal cuts on the photons in~\eqref{acut} we require
\begin{align}
 p_T^{\gamma_{1,2}}>50\ {\rm GeV}~,\quad
 \MET>125\ {\rm GeV}~,\quad
 \Delta\phi(\gamma_{1,2},\,\MET)>0.5~,
\end{align}
as in the ATLAS search~\cite{Aad:2012zza}.
In order to account for the photon reconstruction efficiency of the
detector we multiply by $(0.85)^2$, which is the square of the average
efficiency for a prompt isolated photon with $p_T>20$~GeV at
ATLAS~\cite{Aad:2010sp}. We have checked that the detector simulations with
{\sc Delphes} give similar efficiencies.

For the light sleptons, the efficiencies are decreasing as the emitted
photons become softer. Similarly for large PGLD masses, since the mass
spectrum gets squeezed and the phase space for the photons is
reduced. On the other hand, the slepton-pair production cross sections
drops quickly as the slepton becomes heavier; see
Figure~\ref{fig:xsec}. In total, the visible cross sections for $x=0.9$
are maximal for slepton masses around 175~GeV with a very light PGLD,
where the cross sections reach the observable level, with a maximum
around $1.5$ fb.
Note that ATLAS put a $95\%$ CL upper limit on the visible
cross section at about $1$~fb~\cite{Aad:2012zza}.
In other words,
the region of the parameter space in
the plot \ref{fig:xsec09} leading to a visible cross section larger than $1$~fb
is excluded by the ATLAS search. 
In contrast, for $x=0.5$ the visible cross sections do not reach 1~fb for any of the considered masses and the ATLAS
search does not put any constraint on these models. Clearly, the case $x=0.1$ is even less constrained since the photons are even softer.

Note that in the case where the sleptons are almost degenerate with
the PGLD, i.e.~close to the diagonal, the efficiencies are enhanced.
In this region the neutralino mass is very close to the PGLD mass.
The neutralino decay to the PGLD is then suppressed by the phase space, and the
dominant decay mode is now to the massless GLD (see Figure~\ref{fig:binodecay}).
Hence the emitted photons are significantly harder, and the efficiency is larger.
However, this region also does not reach the experimental bound of $1$~fb
set by ATLAS, neither in the $x=0.5$ case nor in the $x=0.9$ case.

It would be very interesting if ATLAS updated the
inclusive diphoton+$\MET$ search \cite{Aad:2012zza}
 with the full 20 fb$^{-1}$ data set at $\sqrt{s}=8$ TeV.
We checked that the efficiencies only change slightly
 from 7 to 8 TeV, and thus
the plots reported in Figure \ref{fig:eff05} and \ref{fig:eff09} show
that, with a luminosity four times larger, the region of the parameter space
probed by the experiment
will grow considerably in the $x=0.9$ scenario,
and will probably extend also to the $x=0.5$ scenario.

In the $2 \gamma + \MET$ columns of Table \ref{aa8TeV}, assuming the efficiencies of the last
column for the $2 \gamma + \MET$ category in Table~\ref{tab:eff},
we show the expected number of signal events in the existing 20~fb$^{-1}$ of LHC data at $\sqrt{s}=8$~TeV,
for the different benchmark points.
To obtain these numbers we have summed over the three slepton flavors,
taken to be degenerate in mass at $M_{\ell_R}=200$ GeV. The numbers for $K_{G'}=100$ suggest that the diphoton+$\MET$ searches are good probes of the case where the mass splitting between the neutralino and the PGLD are large, i.e.~for $x=0.9$. However, in order to increase the sensitivity to models with smaller mass splittings, we should use different, but complementary, search channels.

\begin{table}
 \center
 \footnotesize
 \begin{tabular}{|rr||r|r|}
 \hline
  \multicolumn{2}{|r||}{$M_{\ell_R}=200$~GeV}
 & $2\gamma+\misset$
 & $2\ell+2\gamma+\misset$ \\
  \multicolumn{2}{|r||}{($K_{G'},\,x,\,M_{G'}$)} & & \\
 \hline\hline
  &(1,\,0.1,\,0)   &  7 & 21\\
 A&(1,\,0.1,\,75)  &  9 & 26\\
  &(1,\,0.1,\,150) & 21 & 22\\
 \hline
  &(1,\,0.5,\,0)   & 11 & 27\\
 B&(1,\,0.5,\,75)  & 13 & 25\\
  &(1,\,0.5,\,150) & 27 & 12\\
 \hline
  &(1,\,0.9,\,0)   & 29 & 8\\
 C&(1,\,0.9,\,75)  & 27 & 3\\
  &(1,\,0.9,\,150) & 44 & 0\\
 \hline
 \end{tabular}
 \quad
 \begin{tabular}{|rr||r|r|}
 \hline
  \multicolumn{2}{|r||}{$M_{\ell_R}=200$~GeV}
 & $2\gamma+\misset$
 & $2\ell+2\gamma+\misset$ \\
  \multicolumn{2}{|r||}{($K_{G'},\,x,\,M_{G'}$)} & & \\
 \hline\hline
  &(100,\,0.1,\,0)   &  7 & 20\\
 A'&(100,\,0.1,\,75) &  1 &  8\\
  &(100,\,0.1,\,150) &  2 &  2\\
 \hline
  &(100,\,0.5,\,0)   & 10 & 27\\
 B'&(100,\,0.5,\,75) &  5 & 20\\
  &(100,\,0.5,\,150) &  0 &  2\\
 \hline
  &(100,\,0.9,\,0)   & 29 &  8\\
 C'&(100,\,0.9,\,75) & 19 &  3\\
  &(100,\,0.9,\,150) &  1 &  0\\
 \hline
 \end{tabular}
 \caption{\small Number of expected signal events with 20~fb$^{-1}$ of
 LHC data at $\sqrt{s}=8$~TeV for the different benchmark points of
 ($K_{G'},\,x,\,M_{G'}$) with $M_{\ell_R}=200$~GeV, where
 two different selection cuts are applied for the $2\gamma+\MET$ and
 $2\ell+2\gamma+\MET$ searches as in the last column, respectively, in
 Table~\ref{tab:eff}.
 \label{aa8TeV}}
\end{table}

Let us now discuss other ways in which our simplified GMSB models can be
probed at the LHC. Since the process in Figure~\ref{fig:stableproc} gives rise to
one opposite-sign same-flavor (OSSF) lepton pair, it makes sense to
select the signal events by requiring the presence of an OSSF lepton
pair in the final state in addition to the two photons and $\MET$. The
benefit of requiring additional particles is that, due to the low
background for the final
state $\ell^+\ell^- +2\gamma+\MET$, it is possible to relax the cuts on
the photon $p_T$ and the $\MET$.

In the last two columns in Table~\ref{tab:eff} the efficiencies are
shown for the $\ell^+\ell^-+2\gamma+\MET$ search. In the column ``min'',
in addition to two photons which pass the minimal cuts in~\eqref{acut},
an OSSF lepton pair with the minimal lepton cuts in~\eqref{lcut} is
required. At the stage of the minimal cuts, i.e. before a selection cut,
the additional lepton requirement reduces the efficiencies with respect
to those in the $2\gamma+\MET$ category.
For large $x$, i.e. for the slepton-neutralino degenerate
scenarios, the leading and sub-leading leptons are too soft to pass the
cuts, as seen in Figures~\ref{fig:e1} and \ref{fig:e2}.
Hence the efficiencies drop significantly.

As a selection cut for $\ell^+\ell^-+2\gamma+\MET$, instead of imposing
high $p_T$ cuts for photons and leptons as well as the missing energy,
we consider a rather soft missing energy cut, $\MET>50$~GeV, on top of
the minimal requirement. All of the benchmark points, except for the
$x=0.9$ case, have much better efficiencies than those in the last
column for $2\gamma+\MET$ in Table~\ref{tab:eff}.
This suggest new search strategies for these scenarios based on final states with
two OSSF leptons plus two photons with softer $p_T$ and $\MET$ cuts. 
The optimization of the kinematical cuts should be done based on a dedicated comparison with background, which is beyond the scope of this paper.

In the $\ell^+ \ell^- + 2 \gamma + \MET$ column of
Table~\ref{aa8TeV}, assuming the efficiencies of the corresponding
column in Table~\ref{tab:eff}, we show the
expected number of signal events
for the existing 20~fb$^{-1}$ of LHC data at $\sqrt{s}=8$~TeV.
Here we only consider selectron and smuon pair production,
discarding stau pair production as a possible production mode. 
Since we expect the irreducible SM background to this
final state to be negligible, the number of expected signal events suggests that 
many of these benchmark points can be probed already with the existing data set.
We also stress that 
the $\ell^+ \ell^- + 2 \gamma + \MET$ search is complementary to the $2 \gamma + \MET$ search, 
since the latter probes the case $x=0.9$, to which the former is less sensitive.

\subsection{The three sector model}\label{3sectorsec}

In this subsection we discuss the signatures of the three sector model in Figure \ref{fig:spectra}, with the relevant process shown in Figure \ref{fig:proc}(b). 
Our analysis will suggest that multi-photon signatures are relevant for GMSB with multiple hidden sectors,
and that they provide new interesting channels that could be searched for both in the
current data set at $\sqrt{s}= 8$ TeV and in the future data set at $13$ TeV. We start by presenting the photon spectra 
and $\MET$ distribution at the partonic level, and then we estimate the number of
expected events at LHC-8TeV with 20 fb$^{-1}$ of integrated luminosity and at LHC-13TeV with 30 fb$^{-1}$.

The mass of the three generations of right-handed sleptons is fixed to be $M_{\ell_R}=200$ GeV as before.
We choose four benchmark points as in Table~\ref{FourBench} for the other mass parameters which highlight the main features of this model.
The spectrum is defined by the masses of the neutralino $M_{\chi}$ and by the two
 PGLD's masses $M_{G''}$ and $M_{G'}$. 
 The benchmark points are chosen with masses separated by at least $50$ GeV
 in order to avoid compressed spectra, leading to very soft photons.
 We label the four benchmark scenarios with three numbers, denoting the particle masses in GeV,
 in the order ($M_{\chi}$,$M_{G''}$,$M_{G'}$).
\begin{table}
\centering
 \begin{tabular}{|c c c c|}
 \hline
$M_{\ell_R}$ & $M_{\chi}$ & $M_{G''}$ & $M_{G'}$\\
 \hline
200 & 150 & 100 & 50 \\
 \hline
200 & 150 & 100 & 0 \\
 \hline
200 & 150 & 50 & 0 \\
 \hline
200 &100 & 50 & 0 \\
 \hline
 \end{tabular}
\caption{\label{FourBench}
The four benchmark points for the three and four photon signals.}
 \end{table}
 The lightest PGLD $\tilde G'$ is collider stable, whereas the decays of the neutralino and the heaviest PGLD
 $\tilde G''$ are prompt, with
 simplified branching ratios $BR(\tilde \chi \to \gamma \tilde G'')=BR(\tilde G'' \to \gamma \tilde G')=100\%$.
 The photons are emitted in two subsequent decays, the first one involving the neutralino and the heaviest PGLD, and
 the second one involving the two PGLD's.

\begin{figure}[ht]
\centering
\subfigure[The $p_T$ distribution of the first photon.]{%
  \includegraphics[width=.45\textwidth]{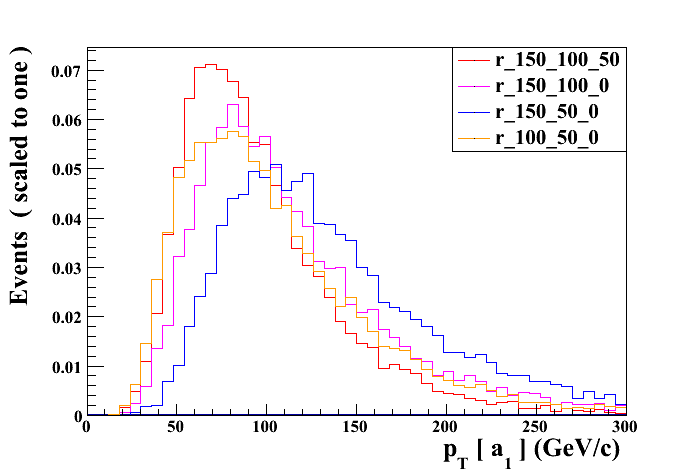}
  \label{fig:ph1}}
\quad
\subfigure[The $p_T$ distribution of the second photon.]{%
  \includegraphics[width=.45\textwidth]{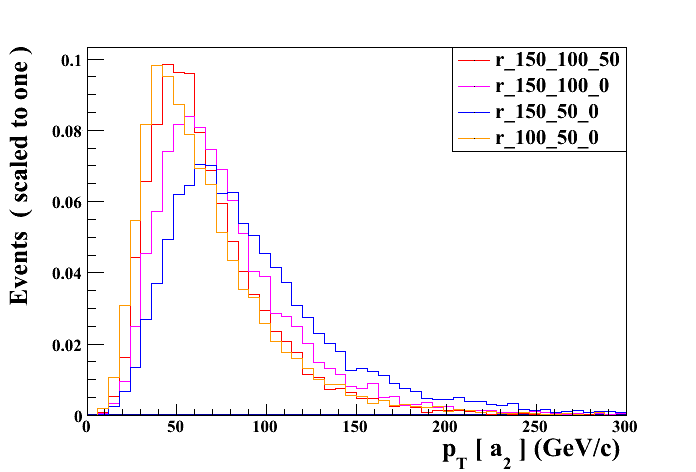}
  \label{fig:ph2}}
\subfigure[The $p_T$ distribution of the third photon.]{%
  \includegraphics[width=.45\textwidth]{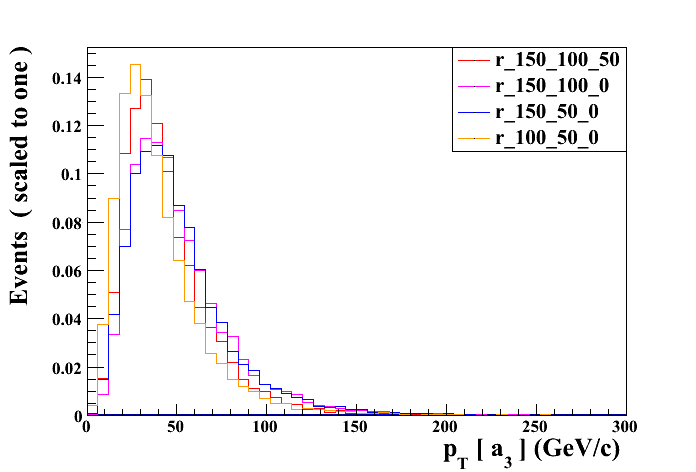}
  \label{fig:ph3}}
\quad
\subfigure[The $p_T$ distribution of the fourth photon.]{%
  \includegraphics[width=.45\textwidth]{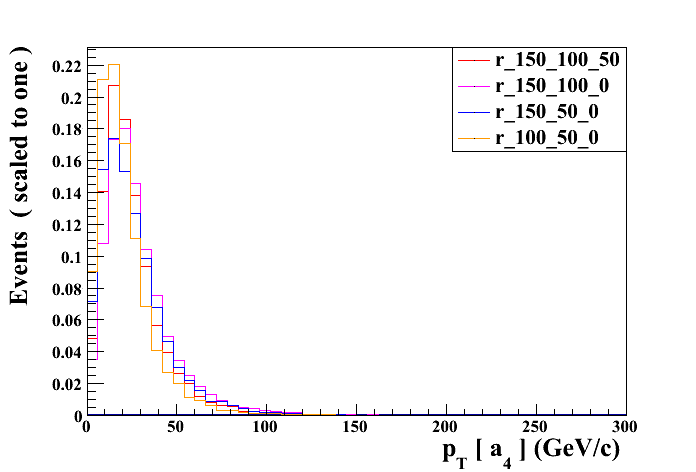}
  \label{fig:ph4}}
 \caption{\small Transverse momentum distributions of the four photons
 for  $pp\to\tilde\ell_R^+\tilde\ell_R^-\to\ell^+\ell^-+ 4 \gamma+\misset$
 at $\sqrt{s}=8$~TeV for 
 the four benchmark points in Table \ref{FourBench}.
  }
\label{fig:4photons}
\end{figure}

\begin{figure}[ht]
\centering
  \includegraphics[width=.45\textwidth]{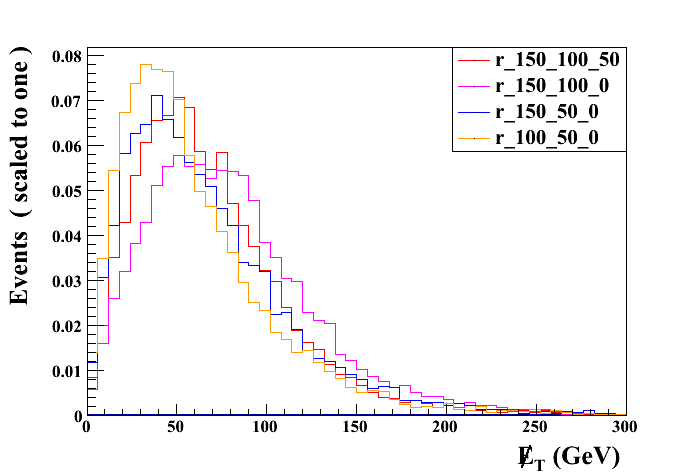}
  \caption{The same as in Fig.\ref{fig:4photons}, but for the missing transverse energy distribution.}
  \label{fig:MET4ph}
\end{figure}

 The normalized $p_T$ distributions for the four photons and $\MET$ are shown in Figures \ref{fig:4photons} and \ref{fig:MET4ph}, respectively. 
%
%
The shape of the distribution of the $p_T$ of the leading photon
differs significantly among the four benchmark points,
as can be seen in Figure \ref{fig:ph1}. The different shapes can be
related to the mass splittings and to the step of the cascade where the hardest photon is emitted,
for each of the benchmark point.
The hardest leading photon is attained for the benchmark point 150-50-0, and is one of the two photons
emitted in the first decay, $\tilde \chi \to \gamma \tilde G''$,
 where the mass difference is large as 100 GeV. The second hardest shape for the leading photon is for the benchmark point 150-100-0;
here the leading photon is emitted in the second decay, $\tilde G'' \to \gamma \tilde G'$, where the mass difference
is 100 GeV. Finally, the two softer cases are the benchmarks where the mass differences
is always 50 GeV. Note that, in the benchmark point 100-50-0, the $p_T$ of the leptons will be maximal, given the
100 GeV mass splitting between the slepton and the neutralino.

In the other photon $p_T$ distributions, the differences among benchmark points are less pronounced
and are correlated with the leading photon distribution shapes we discussed above.
The relevant observation is that the third photon and, even more so, the fourth photon are quite soft, with
mean $p_T$ around $45$ and $25$ GeV respectively.
Hence, imposing a stringent cut on the photon $p_T$, e.g.~$p_T > 50$ GeV, would
strongly suppress the multi-photon signals, leaving only the two leading photons.\footnote{We have checked that the diphoton+$\MET$ ATLAS search \cite{Aad:2012zza} is not constraining the 3 sector benchmark models considered in this section. The main reason is that the $\MET>125$ GeV cut is reducing the efficiency significantly, as can be seen from Figure~\ref{fig:MET4ph}.}

In order to examine the LHC sensitivity to multi-photon final states,
we perform a minimal cut analysis and show the expected number of events,
categorized in different $\MET$ bins.
We select three $\MET$ bins: $(0-50)$ GeV, $(50-100)$ GeV and $(100-\infty)$ GeV.
We distinguish the case in which at least three photons are required in the final state, and the case in which
all the four photons are required.
We consider the following minimal cuts on the identified photons
\begin{equation}
\label{mincuts}
 p_T>20~\text{GeV}~,\quad |\eta|< 2.5~,\quad \Delta R>0.4~.
\end{equation}
These cuts are imposed on the candidate photons, i.e.~on the leading three for the $3 \gamma+\MET$ prospects and on
all four for the $4 \gamma+\MET$ prospect.
The isolation cut is imposed with respect to the other photons and with respect to the leptons.
To take into account detector effects, we estimate the detector efficiency for each photon to be
$85 \%$ \cite{Aad:2010sp}.

\begin{table}[ht]
\centering
\footnotesize
\begin{tabular}{|c|c|c|c|c|c|}
\hline
final state & MET & 150-100-50 & 150-100-0 & 150-50-0 & 100-50-0 \\
\hline
$3 \gamma$ & (0-50) & 32 & 25 & 39 & 43 \\
& (50-100) & 34 & 37 & 32 & 27 \\
 & (100-$\infty$) & 11 & 19 & 14 & 9  \\
 \hline
\hline
final state & MET & 150-100-50 & 150-100-0 & 150-50-0 & 100-50-0 \\
\hline
$4 \gamma$  & (0-50) & 16 & 13 & 19 & 18  \\
& (50-100) & 15 & 19 & 13 & 9  \\
 & (100-$\infty$) & 3.4 & 8.3 & 5.6 & 3.0  \\
 \hline
\end{tabular}
 \caption{\small Number of expected signal events with at least three or four photon final states in the three sector model, using 20~fb$^{-1}$ of data at $\sqrt{s}=8$ TeV, imposing the minimal cuts described in~\eqref{mincuts}.}
 \label{aaa8TeV}
\end{table}

\begin{table}[ht]
\centering
\footnotesize
\begin{tabular}{|c|c|c|c|c|c|}
\hline
final state & MET & 150-100-50 & 150-100-0 & 150-50-0 & 100-50-0 \\
\hline
$3 \gamma$  & (0-50) & 98 & 81 & 120 & 139 \\
& (50-100) & 111  & 114 & 105 & 89 \\
 & (100-$\infty$) & 40 & 67 & 51 & 35  \\
 \hline
\hline
final state & MET & 150-100-50 & 150-100-0 & 150-50-0 & 100-50-0 \\
\hline
$4 \gamma$  & (0-50) & 46  & 41 & 59 & 57  \\
& (50-100) & 47 & 55 & 43 & 29  \\
 & (100-$\infty$) &  14 & 29 & 19 & 12  \\
 \hline
\end{tabular}
  \caption{\small The same as in Table~\ref{aaa8TeV}, but using 30~fb$^{-1}$ of data at $\sqrt{s}=13$ TeV.}
  \label{Table13}
\end{table}

In Table \ref{aaa8TeV} we show the number of expected signal events with at least three or four photons in the final state
for the four benchmark points in Table \ref{FourBench},
with 20fb${}^{-1}$ of data at $\sqrt{s}=8$ TeV, divided in $\MET$ bins.
Clearly the $3 \gamma +\MET$ channel gives rise to more signal events than the $4 \gamma +\MET$ one. The reason is the $p_T > 20$ GeV requirement on the fourth photon, and its $85 \%$ detector efficiency,
which reduce the signal yield in the four photon case.
A cut on $\MET >50$ GeV leaves quite a large number of expected events,
with an efficiency generically larger than $50 \%$.
A more severe cut on the $\MET$ reduces  the signal considerably.
However, even when imposing $\MET > 100$ GeV, the $3 \gamma +\MET$ channel would still produce a significant number of events.

The expected number of signal events in these models should of course be compared with the  SM background for the corresponding final state. The irreducible background for the three and four photon final states are very suppressed in the SM. Instead, the main background
is expected to be the reducible background from misidentified jets,
although the precise estimation should be done carefully.

We have argued that a multi-photon analysis  could lead
to an observation (or very strong constraints) already with the $8$ TeV 20 fb$^{-1}$
data set. Clearly such an analysis would give even stronger results if performed in the next run of LHC. In Table \ref{Table13},  we show the predicted number of events with at least three or four photons
for the same benchmark points as in Table \ref{FourBench}
at LHC $13$ TeV with $30$ fb$^{-1}$.

As a final comment, it is of course possible to consider models with more than three SUSY breaking hidden sectors, and more PGLDs. Even though such models would be analogous to the three sector case, they could in principle give rise to additional prompt decay steps, emitting additional (soft) photons. However, in the LHC searches we propose there is no veto on additional photons and the $p_T$ requirement on the photons are as loose as possible. Therefore, these LHC searches are sensitive to models with any number of hidden sectors.

\vspace{-.3cm}
\section{Conclusions}
\label{Conclusions}

In this paper we discussed how the signatures of standard GMSB are modified in models where SUSY is broken in more than one hidden sector. The first general feature of these models is that the LOSP prefers to decay to a massive PGLD, rather than the nearly massless gravitino, implying that the final state spectrum is softer than in standard GMSB. The second feature is that, in models with more than two hidden sectors, the PGLD can decay promptly to a photon plus another PGLD or a gravitino, implying the possibility of additional \mbox{photons in the final state.}

We focused on the case with a Bino-like neutralino LOSP, decaying to a photon and a PGLD, producing final states involving $2(4)\gamma+\MET$ in the case of two(three) hidden sectors. It is of course possible to consider a different LOSP. For example, if instead the LOSP had been a right-handed slepton, decaying to a lepton and a PGLD, the final state would have been $\ell^{+}\ell^{-}+0(2)\gamma+\MET$, again for two(three) hidden sectors respectively. Note that, in any such scenario, the number of photons in the final state is an indication of the number of hidden sectors.

Concerning the main production mode of SUSY particles, we focused on slepton pair production.
This was our prototypical choice and was motivated by the structure of the soft terms in GMSB.
We have studied in detail the sensitivity of the ATLAS diphoton+$\MET$ search with $4.8$~fb$^{-1}$
at $\sqrt{s}=7$ TeV \cite{Aad:2012zza}
to the scenario with two hidden sectors.
We have shown that a small portion of the parameter space is already excluded by this search,
and that an update of this search, based on the full $20$ fb$^{-1}$ data set at $\sqrt{s}=8$ TeV, would explore a  larger region of the parameter space. However, in order to probe the entire parameter space of these models, new and dedicated searches are needed.

We have proposed inclusive searches in the final states
$({\geqslant}3)\gamma+\MET$ and $\ell^{+}\ell^{-}+({\geqslant}2)\gamma+\MET$.
We showed that, with a cut on \mbox{$\MET>50$ GeV} (and with $p_T>20$ GeV for the photons and leptons), searches in these final states could lead to a discovery (or exclusion) already by using the existing 
LHC data.
The general lesson we draw from our investigation is that GMSB models with multiple hidden sectors can be 
probed by combining inclusive multi-photon searches with searches for photons in association with other final state particles.

Besides slepton pair production, one could envisage other types of electroweak production
and it would be interesting to repeat the study of collider signatures of multiple hidden sector GMSB models
in these cases.
Moreover, in scenarios where some of the colored particles are light enough to be produced at a significant rate, the relevant final states could consist of jets+$4\gamma+\MET$.\footnote{In the case where the lightest stop mass eigenstate is accessible at the LHC, stop pair production can give rise to the final state $t\bar{t}+4\gamma+\MET$.} These models are probably highly constrained by the $S_T$ variable\footnote{The scalar sum of all $p_T$s, including jets (and $\MET$, if above 50~GeV).}, as discussed in~\cite{Evans:2013jna}. Still, it would be interesting to check a few potentially interesting cases given the ease in which these could be excluded.

\subsection*{Acknowledgments}

We would like to thank H.~C.~Cheng, E.~Conte, M.~D'Onofrio, K.~De Causmaecker, B.~Fuks, B.~Heinemann, A.~Hoecker, J.~Keaveney, M.~Spannowsky and S.~Thomas for helpful discussions. We are especially grateful to R.~Argurio and N.~Craig for valuable comments on the draft. This work is supported in part by the Belgian Federal Science Policy Office
through the Interuniversity Attraction Pole P7/37.
The research of G.F. is supported in part by the Swedish Research
Council (Vetenskapsr{\aa}det contract B0508101). A.M. acknowledges funding by the Durham International Junior Research Fellowship. 
K.M. is supported in part by the Strategic Research Program ``High Energy
Physics'' and the Research Council of the Vrije Universiteit Brussel.
The work of C.P. is supported by the Swedish Research Council (VR) under the contract 637-2013-475, by IISN-Belgium (conventions 4.4511.06, 4.4505.86 and 4.4514.08) and by the ``Communaut\'e Fran\c{c}aise de Belgique" through the ARC program and by a ``Mandat d'Impulsion Scientifique" of the F.R.S.-FNRS.

\appendix

\section{The three body decays of the PGLD}

In this appendix we study the three body decays of the PGLD, present the formulae for their widths and show that they are too small to be of any relevance for collider physics.

\subsubsection*{Decay into a pair of fermions}

The effective vertex for the decay of a PGLD into a lighter PGLD and a pair of massless fermions can be derived from the SUSY Lagrangian by integrating out the superpartner of the fermion. Let us denote by $\psi$ the fermion field and by $\tilde\phi$ its scalar superpartner.

The relevant part of the lagrangian is
\beq
     {\mathcal{L}} \supset \sum_i -m^2_{(i)} \tilde\phi^\dagger \tilde\phi + \frac{m^2_{(i)}}{f_i}(\tilde\phi \psi^\dagger\tilde{\eta}_i^\dagger + \tilde\phi^\dagger \psi\tilde{\eta}_i)
     - \frac{m^2_{(i)}}{f^2_i}\psi^\dagger\tilde{\eta}_i^\dagger \psi\tilde{\eta}_i~,
\eeq
where $m_{(i)}$ are the soft masses for $\tilde\phi$ arising from sector $i$.

Integrating out the scalar field $\tilde\phi$, rotating the PGLDs to their mass eigenbasis as in~(\ref{roteta}) and substituting back gives after some simplifications:
\beq
     {\mathcal{L}}_{\psi} = \sum_{ab} K_{ab} \psi \tilde{G}^{(a)} \psi^\dagger \tilde{G}^{(b)\dagger}~, \label{ggpsipsi}
\eeq
where we defined
\beq
      K_{ab} = \sum_{jk} \left(\frac{m^2_{(k)} m^2_{(j)}}{m^2 f_k f_j} - \frac{m^2_{(k)}}{f^2_k} \delta_{jk}\right) V_{ja} V_{kb}~,
\eeq
and, as usual, $f^2 =  \sum_i f^2_i$ and $m^2 =  \sum_i m^2_{(i)}$.

Notice that the true goldstino completely drops out of the Lagrangian (\ref{ggpsipsi}). There are also terms with derivatives obtained from the scalar kinetic term but they are further suppressed by $\partial^2/m^2$. These terms would be the dominant ones for the decay to the true goldstino~\cite{Cheng:2010mw}.

Using (\ref{ggpsipsi}) the decay rate can be easily obtained. Let us
consider the three sector model for definitiveness and define $x \equiv
M_{G'}/M_{G''} < 1$. We find
\beq
    \Gamma(\tilde{G}''\to\tilde{G}'\psi\psi) = \frac{K_{12}^2 M^5_{G''}}{3072\pi^3}(1 - 8x^2 + 8 x^6 - x^8 - 24 x^4 \log x)~. \label{Wll}
\eeq
This is essentially the decay rate of $\tilde{G}''$ into $\tilde{G}'$ and a neutrino pair of a given flavor.
The width for the decay into leptons is obtained by adding the (non-interfering) contributions of $\ell_L$ to those of $\ell_R$ each of them given by the above formula with the appropriate soft parameters. The width for the decay into a quark pair contains an additional color factor $3$.

Consider now the case $f_1 \gg f_2 \gg f_3$ as in the paper. We have
\beq
     K_{12} \simeq \frac{m^2_{(2)} m^2_{(3)}}{m^2 f_2 f_3}~.
\eeq
For the minimal mediation case, with soft parameters scaling like $\frac{\alpha}{4\pi}\frac{f_i}{M_\mathrm{mess}}$, we obtain
\beq
     K_{12} \simeq \frac{\alpha^2}{16 \pi^2}\frac{f_2 f_3}{f^2_1 M^2_\mathrm{mess}}~,
\eeq
For the direct case, where the masses are $\alpha \sqrt{f_i}$, we get
\beq
     K_{12} \simeq \alpha^2 \frac{1}{f_1}~.
\eeq
In either cases it can be easily checked that for the allowed numerical values of the parameters the width is too small to lead to decay inside the detector.

\subsubsection*{Decay into a pair of vector bosons}

We proceed in the same way to analyze the decay of a PGLD into a pair of vector bosons. Here the realistic situation is complicated by the mixing angles arising from the rotation of the vector bosons to their mass eigenstates and by the fact that the lightest neutralino has a mass comparable to that of the heaviest PGLD. Since we are only interested in an order of magnitude estimate, we still
use the effective vertex and, to simplify the notation, simply consider the coupling of the PGLD's to a generic $U(1)$ vector multiplet whose fermionic component is denoted by $\tilde\lambda_\alpha$.

The relevant part of the Lagrangian is now
\beq
     {\mathcal{L}} \supset \sum_i - \frac{M_{(i)}}{2} \tilde\lambda^2 + \frac{i M_{(i)}}{2\sqrt{2} f_i} \tilde\lambda \Fslash \tilde{\eta}_i -
     \frac{M_{(i)}}{16 f_i^2} \tilde{\eta}_i\Fslash\Fslash\tilde{\eta}_i
     + {\rm h.c.}~,
\eeq
where $M_{(i)}$ are the soft masses for $\tilde\lambda$ arising from sector $i$ and we let $M = \sum_i M_{(i)}$.
We defined $\Fslash = \sigma^\mu \bar\sigma^\nu F_{\mu\nu}$.
Note that $\Fslash \Fslash = - 2 F_{\mu\nu}F^{\mu\nu} + i \epsilon^{\mu\nu\rho\lambda}F_{\mu\nu}F_{\rho\lambda}$.

In exactly the same way as with the scalars, integrating out $\tilde\lambda$ yields
\beq
     {\mathcal{L}}_{F} = \sum_{ab} K'_{ab} \tilde{G}^{(a)}
     \Fslash\Fslash \tilde{G}^{(b)} + {\rm h.c.} \label{ggFF}
\eeq
with
\beq
      K'_{ab} = \frac{1}{4}\sum_{jk} \left(\frac{M_{(k)} M_{(j)}}{M f_k f_j} - \frac{M_{(k)}}{f^2_k} \delta_{jk}\right) V_{ja} V_{kb}
\eeq
as before without any dependence on $\tilde{G}$, plus additional terms suppressed by $\partial/M$.

For the three sector model, the decay rate from the above Lagrangian is, identifying the massless vector boson with the photon,
\beq
    \Gamma(\tilde{G}''\to\tilde{G}'\gamma\gamma)  \approx \frac{(K'_{12})^2 M_{G''}^7}{3840 \pi^3}(1 - 15x^2 - 80 x^4 + 80 x^6 + 15 x^8 - x^{10}   - 120 x^4(1 + x^2) \log x)~.
\eeq
In this case we have
\beq
    K'_{12} \simeq \frac{M_{(2)} M_{(3)}}{4 M f_2 f_3}~.
\eeq
For the minimal mediation case, with soft parameters scaling like $\frac{\alpha}{4\pi}\frac{f_i}{M_\mathrm{mess}}$, we obtain
\beq
     K'_{12} \simeq  \frac{\alpha}{16 \pi}\frac{1}{f_1 M_\mathrm{mess}}~.
\eeq
For the direct case, where the masses are $\alpha \sqrt{f_i}$, we get
\beq
     K'_{12} \simeq  \frac{\alpha}{4}\frac{1}{\sqrt{f_1 f_2 f_3}}~.
\eeq
In both cases the width is too small to lead to decays within the detector.

\end{document}